\documentclass[aip,cha, preprint,showpacs,showkeys]{revtex4-1}
\usepackage{amssymb, dcolumn,  graphicx, latexsym, amsfonts, bm}

\begin{document}

\title{Anomalous convective transport of the tokamak edge plasma, caused by the 
inhomogeneous ion cyclotron parametric turbulence}

\author{V. S. Mikhailenko}\email[E-mail:]{vsmikhailenko@pusan.ac.kr}
\affiliation{Plasma Research Center, Pusan National University, Busan 46241, South Korea}
\author{V. V. Mikhailenko}\email[E-mail: ]{vladimir@pusan.ac.kr}
\affiliation{BK21 FOUR Information Technology, Pusan National University, Busan 46241, South Korea}
\author{Hae June Lee}\email[E-mail: ]{haejune@pusan.ac.kr}
\affiliation{Department of Electrical Engineering, Pusan National University, Busan 46241, South 
Korea}
\date{\today}

\begin{abstract}
In this paper, we develop the kinetic and hydrodynamic theories 
of the convective mesoscale flows driven by the spatially inhomogeneous electrostatic 
ion cyclotron parametric microturbulence in the pedestal plasma with a sheared poloidal flow. 
The developed kinetic theory predicts the generation of the sheared poloidal convective flow, 
and of the radial compressed flow with radial flow velocity gradient. The developed 
hydrodynamic theory of the convective flows reveals the radial compressed convective 
flow as the dominant factor in the formation of the steep pedestal density profile 
with density gradient exponentially growing with time. This gradient density growth 
is limited by the formation of the radial oscillating with time ion outflow of pedestal 
plasma to scrape-off layer.
\end{abstract}

\maketitle

 \section{Introduction}\label{sec1}
 
It was found in the fast wave (FW) heating experiments\cite{Perkins, Perkins1} on the National 
Spherical Torus eXperiment (NSTX) that a significant 
part of the FW power loss occurs due to the anomalous convective flow of the
collisionless dense hot plasma from the edge to the cold low density scrape-off layer (SOL) 
plasma. In the SOL, this lost hot plasma propagates mostly along SOL magnetic field 
lines to the divertor regions. It was estimated in Ref.\cite{Mikhailenko}, where the theory 
of the ion cyclotron (IC) electrostatic parametric turbulence, 
driven by the strong inhomogeneous FW electric field in the inhomogeneous plasmas, was developed,
that the origin of this convective flow may be the interactions of ions and electrons  
with radially inhomogeneous  microscale IC and drift turbulence. It was found that the radial 
and poloidal velocities of this flow are proportional to the spatial gradient 
of the spectral intensity of the turbulent electric field. This convective flow has a low flow 
velocity in the plasma core and in the SOL, where the levels of the IC parametric turbulence 
and of the drift turbulence are low, and the plasma parameters are weakly inhomogeneous.  
The edge transport barrier with steep density profile (commonly referred to as the 
pedestal) is, therefore, the most preferable region for the development of the intensive  
convective flows. The transport barrier establishes in the regimes of the enhanced 
confinement (or H-mode regime) which is the basic operational regime of tokamaks. 
This regime is characterized by the suppression of the drift type
instabilities by the poloidal sheared flow, spontaneously developed in the tokamak edge, 
with the flow velocity shearing rate comparable with or above the maximum linear 
growth rate of the suppressed instabilities. Note, that the parametric IC instabilities 
have the growth rate much above the observed velocity shearing rate and are not suppressed 
by this flow. Beginning from the first observation of the H-mode of operation, it was found 
that this mode has been accompanied by periodic bursts of the 
ejected to SOL plasma structures, which are labelled as the edge localized modes 
(ELMs)\cite{Zohm, Leonard} and blobs\cite{D'Ippolito} with considerably larger density 
than the surrounding low density cold SOL plasma. 
It is well recognised now that the non-diffusive transport of the heat and matter outwards 
across the magnetic field, which occurs in addition to the anomalous diffusion, is the universal 
phenomenon\cite{D'Ippolito} experimentally observed in tokamaks. 

It was found in Ref.\cite{Mikhailenko}, that the radial velocity of the convective 
flow is in the direction opposite to the gradient of the spectral intensity of the microscale 
turbulence. In the tokamak plasmas, this convective flow moves outward of pedestal 
to the SOL region. It may be concluded therefore that the dynamical processes in the pedestal 
and near SOL are heavily determined by the convective flows formed in this region by the 
inhomogeneous microturbulence and by interplay of the convective flows with the sheared 
poloidal edge flow of a tokamak plasma, which is the inherent component of the 
pedestal development and sustainment. 

The focus of this paper is the detailed kinetic and hydrodynamic theory of the 
non-diffusive convective flows, driven by the spatially inhomogeneous parametric 
microturbulence in the sheared poloidal edge (basic) flow of a 
tokamak plasma. In Sec. \ref{sec2}, we present a detailed description of the 
two-scale approach to the kinetic theory of the IC parametric instabilities driven by the 
inhomogeneous FW in inhomogeneous plasmas, which is employed in this paper. 
In Sec. \ref{sec3}, we present the theory of the mesoscale convective 
flows driven by the inhomogeneous parametric IC and drift microturbulence in the 
poloidal sheared flow. The derived Vlasov equations for ions and electrons, 
which govern the mesoscale evolution of the plasma, reveal the existence of the radial 
and poloidal convective flows with
radially inhomogeneous flow velocities. Because of the presence of such flows, which 
make a dominant contribution to the 
evolution of inhomogeneous plasma to the equilibrium state, the 
conventional theory of the stability of the steady plasma cannot be applied to this 
problem. In this paper, the problem of the origin of the inhomogeneous convective flows 
and of their temporal evolution is analysed employing the nonmodal approach, which 
was developed in our theory of the temporal evolution and suppression of the 
instabilities of a plasma sheared flow across the magnetic fiels\cite{Mikhailenko1, 
Mikhailenko2, Mikhailenko3}. This approach employs the solution of the initial value problem 
instead of the application of the spectral transform 
in time variable, which is generally applied in the stability theory for the uniform 
plasmas with steady uniform flows. In Sec. \ref{sec3}, we develop the nonmodal theory 
of the convective flows with the radially inhomogeneous flow velocity, which develops 
in the poloidal sheared flow of the plasma edge. The convective flow involves the 
poloidal flow with a radially sheared velocity, and the radial compressed/stretched 
flow with a radially inhomogeneous flow velocity, the effect of which on the plasma 
stability was not considered yet. 

In Sec. \ref{sec4}, we derive, as the moments of the electron and ion mesoscale Vlasov 
equations, the fluid equations for the 
plasma with the sheared and compressed/stretched flows. These equations, which contain the 
continuity, the momentum, and the 
temperature equations for both plasma species, are solved as the initial value problem which
reveal the decisive effect of the compressed radial 
flow on the formation of the plasma outflows. Conclusions are presented in Sec. \ref{sec5}.

\section{Two-scale approach to the theory of the parametric instabilities of the inhomogeneous 
plasma driven by the inhomogeneous fast wave}\label{sec2}

Our theory of the microscale turbulence, driven by the inhomogeneous FW in the inhomogeneous 
plasmas, and the theory of the mesoscale convective flows, generated by the inhomogeneous microscale 
IC and drift turbulence, is based on the Vlasov-Poisson system of equation. In this theory 
we use the approximation of the 
slab geometry where $x, y, z$ directions are viewed as corresponding to the radial, poloidal and 
toroidal directions, respectively, of the toroidal coordinate system. Within this 
approximation the  Vlasov equation employed in our theory for the 
velocity distribution function $F_{\alpha}$ of $\alpha$ plasma 
species ($\alpha=i$ for ions and $\alpha=e$ for electrons) in coordinates 
$\mathbf{r}=\left(x, y, z\right)$, 
\begin{eqnarray}
&\displaystyle \frac{\partial F_{\alpha}\left(\mathbf{v}, \mathbf{r}, t\right)}{\partial
t}+\left. \mathbf{v}\frac{\partial F_{\alpha}\left(\mathbf{v}, \mathbf{r}, t\right)}
{\partial\mathbf{r}}+\frac{e_{\alpha}}{m_{\alpha}}\right(\mathbf{E}_{0x}\left(x\right)
+\mathbf{E}_{1}\left(x, t\right)+\tilde{\mathbf{E}}\left(\mathbf{r}, t\right)
\nonumber 
\\ 
&\displaystyle
\left.+\frac{1}{c}\left[\mathbf{v}\times\left(\mathbf{B}_{0}+
\mathbf{B}_{1}\left(\mathbf{r}, t\right)\right)\right]\right)\frac{\partial
F_{\alpha}\left(\mathbf{v}, \mathbf{r}, t\right)}{\partial\mathbf{v}}=0,
\label{1}
\end{eqnarray}
contains the inhomogeneous radial electric field $\mathbf{E}_{0x}\left(x\right)$, which 
governs the basic poloidal sheared flow, the FW electric field $\mathbf{E}_{1}\left(x, t\right)$, 
the electric field $\tilde{\mathbf{E}}\left(\mathbf{r}, t\right)$ of the self-consistent 
plasma respond on FW, the uniform plasma-confining magnetic field $\mathbf{B}_{0}$ 
directed along the coordinate $z$, and FW magnetic field $\mathbf{B}_{1}\left(\mathbf{r}, t\right)$. 
In the edge layer of the tokamak plasma, the spatial inhomogeneity lengths of 
$\mathbf{E}_{0x}\left(x\right)$ and of FW fields are commensurable with a spatial inhomogeneity 
length of the pedestal plasma density.  This mesoscale spatial inhomogeneity length on order 
of the pedestal width is much less than the wavelength of FW and the inhomogeneity lengths of the 
plasma parameters in the bulk of the tokamak plasma, but is much
larger than the radial wavelength of the IC parametric microturbulence presented 
in Eq. (\ref{1}) by electric field $\tilde{\mathbf{E}}\left(\mathbf{r}, t\right)$. 
This microturbulence develops in the inhomogeneous pedestal plasma by the inhomogeneous FW.
Therefore, electric field $\tilde{\mathbf{E}}\left(\mathbf{r}, t\right)$ of IC microrurbulence 
involves the microscale and mesoscale spatial inhomogenieties. Thus, the consistent theory which 
involves the theory of the IC parametric turbulence of the inhomogeneous plasma, driven by the 
inhomogeneous FW, and the theory of the mesoscale plasma evolution caused by the mesoscale 
inhomogeneities of the IC microturbulence, should include the treatment of plasma evolution on the 
micro- as well as on the mesoscales. We employ for the development of that theory the well known 
two - scale method (simple introduction to the variety versions and applications of this method is 
presented in Ref.\cite{Nayfeh}), which is used for problems in which the solutions depend
simultaneously on widely different scales. Applying this method to our theory, we introduce, 
jointly with variables $\mathbf{r}= \left(x, y, z\right)$ for the microscale 
fast variations, the spatial $X=\varepsilon x$ variable,
where $\varepsilon \ll 1$ is an artificial dimensionless small parameter, which distinguishes 
the slow mesoscale variations along $x$ direction, $\partial/\partial x=\varepsilon\partial/
\partial X$, in the Vlasov equation.

The Vlasov equation (\ref{1}) in two-scales presentation has a form
\begin{eqnarray}
&\displaystyle \frac{\partial F_{\alpha}\left(\mathbf{v}, \mathbf{r}, X, t\right)}{\partial
t}+\mathbf{v}\left(\frac{\partial F_{\alpha}\left(\mathbf{v}, \mathbf{r}, X, t\right)}
{\partial\mathbf{r}}+\varepsilon\frac{\partial F_{\alpha}\left(\mathbf{v}, \mathbf{r}, X, t\right)}
{\partial X}\right)
\nonumber 
\\ 
&\displaystyle
+\frac{e_{\alpha}}{m_{\alpha}}\left(\mathbf{E}_{0x}\left(X\right)
+\mathbf{E}_{1}\left(X, t\right)+\tilde{\mathbf{E}}\left(\mathbf{r}, X, t\right)\right.
\nonumber 
\\ 
&\displaystyle
\left.+\frac{1}{c}\left[\mathbf{v}\times\left(\mathbf{B}_{0}+
\mathbf{B}_{1}\left(X, t\right)\right)\right]\right)\frac{\partial
F_{\alpha}\left(\mathbf{v}, \mathbf{r}, X, t\right)}{\partial\mathbf{v}}=0,
\label{2}
\end{eqnarray}
where FW fields
\begin{eqnarray}
&\displaystyle \mathbf{E}_{1}\left(X, t\right)=
\mathbf{E}_{1x}\left(X\right)\cos \omega_{0}t+ \mathbf{E}_{1y}\left(X\right)\sin \omega_{0}t,
\label{3}
\end{eqnarray}
\begin{eqnarray}
&\displaystyle 
\mathbf{B}_{1}\left(X, t\right)=\frac{c}{\omega_{0}}\frac{dE_{1y}\left(X\right)}{d X}
\cos \omega_{0}t\,\mathbf{e}_{z}.
\label{4}
\end{eqnarray}
are inhomogeneous only on the mesoscale. The electric field $ \tilde{\mathbf{E}}
\left(\mathbf{r}, X, t\right)$, which is inhomogeneous on the micro- and mesoscale, 
is governed by the Poisson equation, 
\begin{eqnarray}
&\displaystyle 
\nabla\cdot \tilde{\mathbf{E}}\left(\mathbf{r}, X, t\right)=
4\pi\sum_{\alpha=i,e} e_{\alpha}\int f_{\alpha}\left(\mathbf{v}, \mathbf{r}, X, t \right)d{\bf v}, 
\label{5}
\end{eqnarray}
in which $f_{\alpha}$ is the fluctuating part of the distribution function 
$F_{\alpha}\left(\mathbf{v}, \mathbf{r}, X, t \right)$, $f_{\alpha}\left(\mathbf{v}, 
\mathbf{r}, X, t \right)=F_{\alpha}\left(\mathbf{v}, \mathbf{r}, X, t \right)-F_{0\alpha}
\left(\mathbf{v}, X, t \right)$. $F_{0\alpha}$ is the equilibrium distribution function of the 
inhomogeneous plasma; this function is governed by Eq. (\ref{2}) with 
$\tilde{\mathbf{E}}\left(\mathbf{r}, X, t\right)=0$.

The central point in the theory of the IC parametric instabilities driven by 
the spatially inhomogeneous FW in the inhomogeneous plasma without sheared flow, developed Ref.
\cite{Mikhailenko}, is the transformation of the velocity $\mathbf{v}$, mesoscale 
coordinate $X$ and microscale coordinates $\mathbf{r}= \left(x, y, z\right)$ 
in Eq. (\ref{1}) to the velocity $\mathbf{v}_{i}$, the mesoscale coordinate $X_{i}$ 
and the microscale coordinates $\mathbf{r}_{i}$, determined in the frame of reverences 
which moves with velocity 
$\mathbf{V}_{i}\left(X_{i}, t\right)$ equal to the 
velocity of an ion in the plasma-confining magnetic field $\mathbf{B}_{0}$,  and FW fields
relative to the laboratory frame.
The transformation $\left(\mathbf{v}, X,\mathbf{r}, t\right)\rightarrow \left(\mathbf{v}_{i}, X_{i},
\mathbf{r}_{i}, t\right)$ is performed by employing the relation 
\begin{eqnarray} 
&\displaystyle
\mathbf{v}=\mathbf{v}_{i}+\mathbf{V}_{i}\left(X_{i}, t\right),
\label{6}
\end{eqnarray}
and the inverse relation
\begin{eqnarray}
&\displaystyle
\mathbf{v}_{i}=\mathbf{v}-\mathbf{V}\left(X, t\right); 
\label{7}
\end{eqnarray}
the relation
\begin{eqnarray}
&\displaystyle
X=X_{i}+R_{ix}\left(X_{i}, t\right)=X_{i}+\int\limits^{t}_{t_{0}}V_{ix}
\left(X_{i}, t_{1}\right)dt_{1},
\label{8}
\end{eqnarray}
and the inverse relation,
\begin{eqnarray} 
&\displaystyle
X_{i}= X-\int\limits^{t}_{t_{0}}V_{x}\left(X, t_{1}\right)dt_{1}. 
\label{9}
\end{eqnarray}
The  velocity $\mathbf{V}_{i}\left(X_{i}, t\right)$ is determined by the 
equation\cite{Mikhailenko} 
\begin{eqnarray}
&\displaystyle 
\frac{d\mathbf{V}_{i}}{dt}
=\frac{e_{i}}{m_{\alpha}}\left(\mathbf{E}_{1}\left(X_{i}+
R_{ix}\left(X_{i}, t\right), t\right)
+\frac{1}{c}\left[\mathbf{V}_{i}\times\mathbf{B}_{0}\right]\right.
\nonumber 
\\
&\displaystyle
\left.+\frac{1}{c}\left[\mathbf{V}_{i}\times\mathbf{B}_{1}\left(X_{i}+R_{ix}
\left(X_{i}, t\right), t\right)\right]\right),
\label{10}
\end{eqnarray}
with initial value $\mathbf{V}_{i}\left(X_{i}, t=t_{0}\right)=0$. In this equation, the variable 
$X_{i}$, which determines the spatial inhomogeneity of the plasma 
and FW field, enters to Eq. (\ref{10}) as a parameter. 
It was derived in Ref.\cite{Mikhailenko} that when the amplitude of the ion 
displacement $\left|R_{ix}\left(X_{i}, t\right)\right|$ in FW field is 
much less than the spatial scale length $L_{E}$ of the radial inhomogeneity of the FW field, 
the solution to Eq. (\ref{10}) may be derived 
in a form of a power series expansion in powers of $R_{ix}/L_{E}$. It was found \cite{Mikhailenko} 
that when $\omega_{0}-\omega_{ci}\sim \omega_{ci}$, where $\omega_{ci}$ is the ion cyclotron 
frequency, the terms on the order of $O\left(R_{ix}/L_{E}\right)$ of this expansions with FW for 
pedestal parameters are negligibly small and may 
be neglected. It was derived in Ref.\cite{Mikhailenko} that with variables 
$\mathbf{v}_{i}$, $X_{i}$, and $\mathbf{r}_{i}$, where
\begin{eqnarray}
&\displaystyle
\mathbf{r}=\mathbf{r}_{i}+\mathbf{R}_{i}\left(X_{i}, t\right)
=\mathbf{r}_{i}+
\int\limits^{t}_{t_{0}}\mathbf{V}_{i}\left(X_{i}, t\right)dt_{1},
\label{11}
\end{eqnarray}
FW electric field is presented in the Vlasov equation (\ref{2}) for the microscale perturbations 
only in terms of the order of $|R_{ix}/L_{E}|\ll 1$ which may be neglected. Therefore the  
Vlasov equation for $F_{i}\left(\mathbf{v}_{i}, \mathbf{r}_{i}, X_{i}, t\right)$ 
with great accuracy has a form as for a steady 
plasma in the uniform magnetic field $\mathbf{B}_{0}$ without FW fields, and contains the 
variable $X_{i}$ as a parameter, i. e.
\begin{eqnarray}
&\displaystyle 
\frac{\partial F_{i}\left(\mathbf{v}_{i}, \mathbf{r}_{i}, X_{i}, t\right)}
{\partial t}+ \mathbf{v}_{i}\frac{\partial F_{i}} {\partial
\mathbf{r}_{i}}+\frac{e_{i}}{m_{i}c}\left[\mathbf{v}_{i}\times\mathbf{B}_{0}\right]
\frac{\partial F_{i}}{\partial\mathbf{v}_{i}}
\nonumber 
\\ 
&\displaystyle
+\frac{e_{i}}{m_{i}}\tilde{\mathbf{E}}_{i}\left(\mathbf{r}_{i}, X_{i}, t\right)
\frac{\partial F_{i}\left(\mathbf{v}_{i}, \mathbf{r}_{i}, X_{i}, 
t\right)}{\partial \mathbf{v}_{i}} =0.
\label{12}
\end{eqnarray}
In fact, this equation is the result of the quantitative justification\cite{Mikhailenko} of 
the application of the local approximation to the theory of the parametric instabilities 
driven by the inhomogeneous FW field in the pedestal plasma. 
This result is completely applicable to the case of FW in the poloidal sheared 
flow of a tokamak plasma. In this case, velocity $\mathbf{V}_{i}$ is determined by Eq. (\ref{10}) 
with the electric field $\mathbf{E}_{1}\left(X_{i}+
R_{ix}\left(X_{i}, t\right), t\right)+\mathbf{E}_{0x}\left(X_{i}+
R_{ix}\left(X_{i}, t\right)\right)$.  

The equation similar to Eq. (\ref{12}) is derived for the Vlasov equation 
for the electron distribution function 
$F_{e}\left(\mathbf{v}_{e}, \mathbf{r}_{e}, X_{e}, t\right)$, where $\mathbf{v}_{e}$ 
and $\mathbf{r}_{e}$ are the velocity and the 
position vector of electron in the frame of references which moves with velocity 
$\mathbf{V}_{e}\left(X_{e}, t\right)$ determined by Eq. 
(\ref{10}) with ion species subscript $i$ changed on the electron subscript $e$. 
It follows from Eq. (\ref{12}) that the Vlasov equation for the perturbation $f_{i}\left(\mathbf{v}_{i},\mathbf{r}_{i}, X_{i}, t \right)$ 
of the ion distribution function $F_{i0}\left(\mathbf{v}_{i}, X_{i}\right)$ and the similar equation for the perturbation 
$f_{e}\left(\mathbf{v}_{e},\mathbf{r}_{e}, X_{e}, t \right)$ of the 
electron distribution function $F_{e0}\left(\mathbf{v}_{e}, X_{e}\right)$ are the same as for a plasma without FW field,
\begin{eqnarray}
&\displaystyle 
\frac{\partial f_{i}\left(\mathbf{v}_{i},\mathbf{r}_{i}, X_{i}, t \right)}{\partial t}+\mathbf{v}_{i}\frac{\partial f_{i}}{\partial\mathbf{r}_{i}}
+\frac{e_{i}}{m_{i}c}\left[\mathbf{v}_{i}\times\mathbf{B}_{0}\right]\frac{\partial f_{i}}{\partial\mathbf{v}_{i}}
\nonumber 
\\ 
&\displaystyle
+\frac{e_{i}}{m_{i}}\tilde{\mathbf{E}}_{i}\left(\mathbf{r}_{i}, X_{i}, t\right)\frac{\partial
f_{i}}{\partial\mathbf{v}_{i}} =-\frac{e_{i}}{m_{i}}\tilde{\mathbf{E}}_{i}\left(\mathbf{r}_{i}, X_{i}, t\right)\frac{\partial 
F_{i0}\left(\mathbf{v}_{i}, X_{i}\right)}{\partial\mathbf{v}_{i}},
\label{13}
\\
&\displaystyle 
\frac{\partial f_{e}\left(\mathbf{v}_{e},\mathbf{r}_{e}, X_{e}, t \right)}{\partial t}+\mathbf{v}_{e}\frac{\partial f_{e}}{\partial\mathbf{r}_{e}}
+\frac{e}{m_{e}c}\left[\mathbf{v}_{e}\times\mathbf{B}_{0}\right]\frac{\partial f_{e}}{\partial
\mathbf{v}_{e}}
\nonumber 
\\ 
&\displaystyle
+\frac{e}{m_{e}}\tilde{\mathbf{E}}_{e}\left(\mathbf{r}_{e}, X_{e}, t\right)\frac{\partial
f_{e}}{\partial\mathbf{v}_{e}}=-\frac{e}{m_{e}}\tilde{\mathbf{E}}_{e}
\left(\mathbf{r}_{e}, X_{e}, t\right)\frac{\partial F_{e0}\left(\mathbf{v}_{e}, X_{e}\right)}
{\partial\mathbf{v}_{e}}.
\label{14}
\end{eqnarray}
The mesoscale spatial variables $X_{i}$ and $X_{e}$ are introduced to Eqs. (\ref{13}), (\ref{14})
by the spatial inhomogeneity of $F_{i0}\left(\mathbf{v}_{i}, X_{i}\right)$ and 
$F_{e0}\left(\mathbf{v}_{e}, X_{e}\right)$ functions and present in these equations as a parameters. 
Thus, the solutions to $f_{i}$ and $f_{e}$  also depend on $X_{i}$ and $X_{e}$. 
The solutions to Eqs. (\ref{13}) and (\ref{14}) were derived in Ref.\cite{Mikhailenko}, 
where the local approximation to the theory of the parametric 
IC microinstabilities driven by the inhomogeneous FW in the inhomogeneous plasma was developed.
These results for $f_{i}$ and $f_{e}$, as well as ones for the perturbed ion
and electron densities, are applicable to the present investigations of the IC 
microinstabilities in the poloidal sheared flow. 

The Vlasov equations (\ref{13}) and (\ref{14}) for $f_{i}\left(\mathbf{v}_{i}, 
\mathbf{r}_{i}, X_{i}, t\right)$ and $f_{e}\left(\mathbf{v}_{e}, \mathbf{r}_{e}, 
X_{e}, t\right)$, and 
the Poisson equation 
\begin{eqnarray}
&\displaystyle 
\nabla\cdot \tilde{\mathbf{E}}\left(\mathbf{r}_{i}, X_{i}, t\right)=
4\pi\sum_{\alpha=i,e} e_{\alpha}\int f_{\alpha}\left(\mathbf{v}_{\alpha}, 
\mathbf{r}_{\alpha}, X_{\alpha}, t \right)d\mathbf{v}_{\alpha}, 
\label{15}
\end{eqnarray}
for the electric field $\tilde{\mathbf{E}}_{i}
\left(\mathbf{r}_{i}, X_{i}, t\right)= -\frac{\partial\varphi_{i}
\left(\mathbf{r}_{i}, X_{i}, t\right)} {\partial \mathbf{r}_{i}}$ 
compose the system of equation for the investigation of the microscale IC 
parametric instabilities of a plasma sheared flow in the inhomogeneous FW field, 
and the nonlinear evolution and saturation these instabilities. In the saturation state, 
the electric field $\tilde{\mathbf{E}}$ of the electrostatic 
parametric microturbulence is determined in the ion frame in the form
\begin{eqnarray}
&\displaystyle 
\tilde{\mathbf{E}}_{i}\left(\mathbf{r}_{i}, X_{i}, t\right)=\int d\mathbf{k}
\tilde{\mathbf{E}}\left(\mathbf{k}, X_{i}\right)
e^{-i\omega\left(\mathbf{k}, X_{i}\right)t+ i\mathbf{k}\mathbf{r}_{i}+i\theta
\left(\mathbf{k}\right)}
\nonumber\\ 
& \displaystyle
=-i\int d\mathbf{k}\mathbf{k}\varphi\left(\mathbf{k}, X_{i}\right)
e^{-i\omega\left(\mathbf{k}, X_{i}\right)t+ i\mathbf{k}\mathbf{r}_{i}+i\theta
\left(\mathbf{k}\right)},
\label{16}
\end{eqnarray}
where the integration over $\mathbf{k}$ is performed over wave numbers of the 
linearly unstable IC perturbations, and $\theta\left(\mathbf{k}\right)$ 
is their initial phase. In the electron 
frame, oscillating relative to the ion frame, this electric field is determined by Eq. 
(\ref{16}) with species subscript $i$ changed on $e$. The electric field 
$\tilde{\mathbf{E}}_{e}\left(\mathbf{r}_{e}, X_{e}, t\right)$ in variables 
$\mathbf{r}_{i}, X_{i}$ has a form
\begin{eqnarray}
&\displaystyle 
\tilde{\mathbf{E}}_{e}\left(\mathbf{r}_{e}, X_{e}, t\right)=\int d\mathbf{k}
\tilde{\mathbf{E}}_{e}\left(\mathbf{k}, X_{e}\right)
e^{-i\omega\left(\mathbf{k}, X_{e}\right)t+ i\mathbf{k}\mathbf{r}_{e}+i\theta
\left(\mathbf{k}\right)}
\nonumber\\ 
& \displaystyle
=\int d\mathbf{k}\tilde{\mathbf{E}_{i}}\left(\mathbf{k}, X_{i}\right)
e^{-i\omega\left(\mathbf{k}, X_{i}\right)t + i\mathbf{k}\mathbf{r}_{i}+i\mathbf{k}
\left(\mathbf{r}_{e}-\mathbf{r}_{i}\right)
+i\theta\left(\mathbf{k}\right)}
\nonumber\\ 
& \displaystyle
=\int d\mathbf{k}\tilde{\mathbf{E}}_{i}\left(\mathbf{k}, X_{i}\right)
\sum\limits_{p=-\infty}^{\infty}J_{p}\left(a_{ie}\right)e^{i\mathbf{k}\mathbf{r}_{i}
-i\Omega_{p}\left(\mathbf{k}, X_{i}\right)t-ip\delta_{ie}\left(\mathbf{k}, X_{i}\right)
+i\theta\left(\mathbf{k}\right)},
\label{17}
\end{eqnarray}
where 
\begin{eqnarray}
&\displaystyle 
\Omega_{p}\left(\mathbf{k}, X_{i}\right)=\omega\left(\mathbf{k}, X_{i}\right)
+p\omega_{0},
\label{18}
\end{eqnarray}
and $J_{p}\left(a_{ie}\right)$ is the first kind Bessel 
function of order $p$ with argument $a_{ie}$. Parameters $a_{ie}\sim k\xi_{ie}$, 
where $\xi_{ie}$ is the amplitude of the relative displacement of electrons 
relative to ions in FW field, and $\delta_{ie}$ were determined in Ref.\cite{Mikhailenko}. 

The theory of the parametric microturbulence, developed in details in 
Ref.\cite{Mikhailenko} for the case $V'_{0}=0$, 
is completely applicable to the tokamak poloidal sheared flows, where always the velocity 
shearing rate $V'_{0}=-cE'_{0}/B_{0}$ is much less than IC frequency 
$\omega_{ci}$ and is less than the growth rate of the IC parametric instabilities. 
However, the poloidal sheared flow with the velocity shearing rate commensurable with 
velocity gradients of the inhomogeneous mesoscale convective flows strongly affects 
their structure and temporal evolution. 

\section{Mesoscale convective flows in the poloidally sheared plasma flow}\label{sec3}

On the nonlinear stage of the IC parametric microturbulence evolution at time above 
the inverse growth rate of the IC instabilities, 
$t\gg \gamma^{-1}\left(\mathbf{k}\right)> |\omega^{-1}\left(\mathbf{k}\right)|$, electric field 
(\ref{16}) becomes the random function of the initial phase $\theta
\left(\mathbf{k}\right)$. The motion of ions and electrons in this field has a form 
of the random scattering of particles by the turbulent electric field and mimics to the 
thermal motion of particles. Here we consider the average effect of the mesoscale  
inhomogeneity of the microscale IC turbulence on the mesoscale evolution of the ion 
and electron distribution functions. For this goal, we transform velocity 
$\mathbf{v}_{i}$ and mesoscale $X_{i}$, $Y_{i}$ and microscale 
$\mathbf{r}_{i}$ spatial variables to new velocity $\tilde{\mathbf{v}}_{i}$ 
and coordinates $\tilde{X}_{i}$, $\tilde{Y}_{i}$, $\tilde{\mathbf{r}}_{i}$, 
in which the thermal ion motion and the ion motion in the turbulent electric field 
$\tilde{\mathbf{E}}_{i}$ are separated. The transformation $\left(\mathbf{v}_{i}, 
X_{i}, Y_{i}, \mathbf{r}_{i}, t\right) \rightarrow \left(\tilde{\mathbf{v}}_{i}, \tilde{X}_{i}, 
\tilde{Y}_{i}, \tilde{\mathbf{r}}_{i}, t\right)$ is determined by the relations 
\begin{eqnarray}
&\displaystyle
\tilde{\mathbf{v}}_{i}=\mathbf{v}_{i}-\tilde{\mathbf{V}}_{i}\left(\mathbf{r}_{i}, 
X_{i}, t\right), 
\label{19}
\end{eqnarray}
\begin{eqnarray}
&\displaystyle
\tilde{X}_{i}=X_{i}-\int\limits^{t}_{t_{0}}\tilde{V}_{ix}\left(\mathbf{r}_{i}, X_{i}, t_{1}
\right)dt_{1}, 
\label{20}
\end{eqnarray}
\begin{eqnarray}
&\displaystyle
\tilde{Y}_{i}=Y_{i}-\int\limits^{t}_{t_{0}}\tilde{V}_{iy}\left(\mathbf{r}_{i}, 
X_{i}, t_{1}\right)dt_{1}, 
\label{21}
\end{eqnarray}
and
\begin{eqnarray}
&\displaystyle
\tilde{\mathbf{r}}_{i}=\mathbf{r}_{i}-\int\limits^{t}_{t_{0}}\tilde{\mathbf{V}}_{i}
\left(\mathbf{r}_{i}, X_{i}, t_{1}\right)dt_{1}, 
\label{22}
\end{eqnarray}
or by their inverse,
\begin{eqnarray}
&\displaystyle
\mathbf{v}_{i}=\tilde{\mathbf{v}}_{i}+\tilde{\mathbf{U}}_{i}
\left(\tilde{\mathbf{r}}_{i}, \tilde{X}_{i}, t\right), 
\label{23}
\end{eqnarray}
\begin{eqnarray}
&\displaystyle
X_{i}=\tilde{X}_{i}+\int\limits^{t}_{t_{0}}
\tilde{U}_{ix}\left(\tilde{\mathbf{r}}_{i}, \tilde{X}_{i}, t_{1}\right)dt_{1}, 
\label{24}
\end{eqnarray}
\begin{eqnarray}
&\displaystyle
Y_{i}=Y-V'_{0}Xt =\tilde{Y}_{i}+\int\limits^{t}_{t_{0}}
\tilde{U}_{iy}\left(\tilde{\mathbf{r}}_{i}, \tilde{X}_{i}, t_{1}\right)dt_{1}, 
\label{25}
\end{eqnarray}
and
\begin{eqnarray}
&\displaystyle
\mathbf{r}_{i}=\tilde{\mathbf{r}}_{i}+\tilde{\mathbf{R}}_{i}\left(\tilde{\mathbf{r}}_{i}, \tilde{X}
_{i}, t\right)=\tilde{\mathbf{r}}_{i}+\int\limits^{t}_{t_{0}}
\tilde{\mathbf{U}}_{i}\left(\tilde{\mathbf{r}}_{i}, \tilde{X}_{i}, t_{1}\right)dt_{1}. 
\label{26}
\end{eqnarray}
The velocity $\tilde{\mathbf{V}}_{i}\left(\mathbf{r}_{i}, X_{i}, t\right)$ 
is determined by the Euler equation  
\begin{eqnarray}
&\displaystyle 
\frac{\partial \tilde{\mathbf{V}}_{i}}{\partial t}+\tilde{V}_{ix}\frac{\partial 
\tilde{\mathbf{V}}_{i}\left(\mathbf{r}_{i}, X_{i}, t\right)}{\partial {X}_{i}}
=\frac{e_{i}}{m_{i}}\left(\tilde{\mathbf{E}}_{i}\left(\mathbf{r}_{i}, X_{i}, t\right)
+\frac{1}{c}\left[\tilde{\mathbf{V}}_{i}\times\mathbf{B}_{0}\right]\right)
\label{27}
\end{eqnarray}
as the velocity of an ion in the electric field $\tilde{\mathbf{E}}_{i}
\left(\mathbf{r}_{i}, X_{i}, t\right)$ of the IC parametric turbulence, 
where variables $\mathbf{r}_{i}, X_{i}$ are determined in the frame of references 
which moves with the velocity of an ion in the field of the fast wave in a poloidal 
sheared flow. In variables $\left(\tilde{X}_{i}, t\right)$, the convective 
nonlinear part $\tilde{V}_{ix}\frac{\partial }{\partial {X}_{i}}$ of the operator 
$\frac{\partial}{\partial t}+\tilde{V}_{ix}\frac{\partial }{\partial 
{X}_{i}}$ of Eq. (\ref{27}) vanishes and this operator is transformed to the linear one, $
\frac{\partial}{\partial t}$. Then, Eq. (\ref{27}) becomes the ordinary differential 
equation 
\begin{eqnarray}
&\displaystyle 
\frac{d}{dt}\tilde{\mathbf{U}}_{i}\left(\tilde{\mathbf{r}}_{i}, \tilde{X}_{i}, t\right)
\nonumber\\ 
&\displaystyle
=\frac{e_{i}}{m_{i}}\left(\tilde{\mathbf{E}}_{i}\left(\tilde{\mathbf{r}}_{i}
+\tilde{\mathbf{R}}_{i}\left(\tilde{\mathbf{r}}_{i}, \tilde{X}_{i}, t\right), \tilde{X}_{i}
+\tilde{R}_{ix}\left(\tilde{\mathbf{r}}_{i}, \tilde{X}_{i}, t\right), t\right)+\frac{1}{c}
\left[\tilde{\mathbf{U}}_{i}\left(\tilde{\mathbf{r}}_{i}, \tilde{X}_{i}, t\right)
\times\mathbf{B}_{0}\right]\right).
\label{28}
\end{eqnarray}
for $\tilde{\mathbf{U}}_{i}\left(\tilde{\mathbf{r}}_{i}, \tilde{X}_{i}, t\right)
= \tilde{\mathbf{V}}_{i}\left(\mathbf{r}_{i}, X_{i}, t\right)$, where 
$\mathbf{r}_{i}, X_{i}$ are determined as a functions of $\tilde{\mathbf{r}}_{i}, 
\tilde{X}_{i}, t$ by Eqs. (\ref{26}) and (\ref{24}). The spatial 
micro- and mesoscale variables $\tilde{\mathbf{r}}_{i}$ and $\tilde{X}_{i}$ 
enter to Eq. (\ref{28}) as parameters. Equation (\ref{28}), as well 
as Eq. (\ref{10}), reveals as a hybrid of Eulerian - Lagrangian description of the ion 
dynamics, in which the ion velocity and the ion displacement 
associated with the waves are defined as in the Eulerian dynamics by the functions of 
$\tilde{X}_{i}$ and $t$ and not primarily as a function of the initial position of ion as 
in a purely Lagrangian description. At the same time, the nonlinear convective derivative 
is absent in Eqs. (\ref{28}) and (\ref{10}). 

Using the expansion for the electric field $\tilde{\mathbf{E}}_{i}$ in Eq. (\ref{28}),
\begin{eqnarray}
&\displaystyle 
\tilde{\mathbf{E}}_{i}\left(\tilde{\mathbf{r}}_{i}
+\tilde{\mathbf{R}}_{i}\left(\tilde{\mathbf{r}}_{i}, \tilde{X}_{i}, t\right), \tilde{X}_{i}
+\tilde{R}_{ix}\left(\mathbf{\tilde{r}}_{i}, \tilde{X}_{i}, t\right), t\right)\approx 
\tilde{\mathbf{E}}_{i}\left(\tilde{\mathbf{r}}_{i}
+\tilde{\mathbf{R}}_{i}\left(\tilde{\mathbf{r}}_{i}, \tilde{X}_{i}, t\right), 
\tilde{X}_{i}, t\right)
\nonumber 
\\ &\displaystyle
+\tilde{R}_{ix}\left(\tilde{\mathbf{r}}_{i}, 
\tilde{X}_{i}, t\right)\frac{\partial }
{\partial \tilde{X}_{i}}\tilde{\mathbf{E}}_{i}\left(\tilde{\mathbf{r}}_{i}+\tilde{\mathbf{R}}_{i}
\left(\tilde{\mathbf{r}}_{i}, \tilde{X}_{i}, t\right), \tilde{X}_{i}, t\right),
\label{29}
\end{eqnarray}
which is valid for the small displacement, $\left|\tilde{R}_{ix}\left(\tilde{\mathbf{r}}_{i}+
\tilde{\mathbf{R}}_{i}, \tilde{X}_{i}, t\right)\right| \ll L_{\tilde{E}}$, of an ion in the 
inhomogeneous electric field $\tilde{\mathbf{E}}_{i}$, the solution to Eq. (\ref{28}) 
with the initial value 
$\tilde{\mathbf{U}}_{i}\left(\mathbf{\tilde{r}}_{i}, \tilde{X}_{i}, t_{0} \right)=0$
is easily derived and may be presented in the form
\begin{widetext}
\begin{eqnarray}
&\displaystyle 
\tilde{U}_{ix}\left(\tilde{\mathbf{r}}_{i}, \tilde{X}_{i}, t\right) \approx
\tilde{U}^{(0)}_{ix}\left(\tilde{\mathbf{r}}_{i}, \tilde{X}_{i}, t\right)+
\tilde{U}^{(1)}_{ix}\left(\tilde{\mathbf{r}}_{i}, \tilde{X}_{i}, t\right),
\label{30}
\end{eqnarray}
and
\begin{eqnarray}
&\displaystyle 
\tilde{U}_{iy}\left(\tilde{\mathbf{r}}_{i}, \tilde{X}_{i}, t\right) \approx
\tilde{U}^{(0)}_{iy}\left(\tilde{\mathbf{r}}_{i}, \tilde{X}_{i}, t\right)+
\tilde{U}^{(1)}_{iy}\left(\tilde{\mathbf{r}}_{i}, \tilde{X}_{i}, t\right).
\label{31}
\end{eqnarray}
In Eqs. (\ref{30}), (\ref{31})
\begin{eqnarray}
&\displaystyle 
\tilde{U}^{(0)}_{ix}\left(\tilde{\mathbf{r}}_{i}, \tilde{X}_{i}, t\right) 
= \frac{e_{i}}{m_{i}}\int\limits ^{t}_{t_{0}}dt_{1}\left[\tilde{E}_{x}
\left(\tilde{\mathbf{r}}_{i}+\tilde{\mathbf{R}}_{i}, \tilde{X}_{i}, t_{1}\right)\cos \omega_{ci}
\left(t-t_{1}\right)\right. 
\nonumber 
\\ &\displaystyle
\left.+ \tilde{E}_{y}\left(\tilde{\mathbf{r}}_{i}+\tilde{\mathbf{R}}_{i}, \tilde{X}_{i}, t_{1}\right)
\sin \omega_{ci}\left(t-t_{1}\right) \right], 
\label{32}
\end{eqnarray}
\begin{eqnarray}
&\displaystyle 
\tilde{U}^{(0)}_{iy}\left(\tilde{\mathbf{r}}_{i}, \tilde{X}_{i}, t\right)
= \frac{e_{i}}{m_{i}}\int\limits ^{t}_{t_{0}}dt_{1}\left[-\tilde{E}_{x}
\left(\tilde{\mathbf{r}}_{i}+\tilde{\mathbf{R}}_{i}, \tilde{X}_{i}, t_{1}\right)\sin\omega_{ci}
\left(t-t_{1}\right)\right. 
\nonumber 
\\ &\displaystyle
\left.+ \tilde{E}_{y}
\left(\tilde{\mathbf{r}}_{i}+\tilde{\mathbf{R}}_{i}, \tilde{X}_{i}, t_{1}\right)\cos\omega_{ci}
\left(t-t_{1}\right) \right], 
\label{33}
\end{eqnarray}
\begin{eqnarray}
&\displaystyle 
\tilde{U}^{(1)}_{ix}\left(\tilde{\mathbf{r}}_{i}, \tilde{X}_{i}, t\right) 
\nonumber 
\\ &\displaystyle = \frac{e_{i}}{m_{i}}\int\limits ^{t}_{t_{0}}dt_{1}\left[\frac{\partial }
{\partial \tilde{X}_{i}}\left(\tilde{E}_{x}
\left(\tilde{\mathbf{r}}_{i}+\tilde{\mathbf{R}}_{i}, \tilde{X}_{i}, t_{1}\right)\right)
\cos \omega_{ci}\left(t-t_{1}\right)
\int\limits^{t_{1}}_{t_{0}}dt_{2}\tilde{U}^{(0)}_{ix}\left(\tilde{\mathbf{r}}_{i}
+\tilde{\mathbf{R}}_{i}, \tilde{X}_{i}, t_{2}
\right) \right. 
\nonumber 
\\ &\displaystyle
\left.+ \frac{\partial}{\partial \tilde{X}_{i}}\left(\tilde{E}_{y}
\left(\tilde{\mathbf{r}}_{i}+\tilde{\mathbf{R}}_{i}, \tilde{X}_{i}, t_{1}\right)\right)
\sin \omega_{ci}\left(t-t_{1}\right)\int\limits^{t_{1}}_{t_{0}}dt_{2}
\tilde{U}^{(0)}_{ix}\left(\tilde{\mathbf{r}}_{i}+\tilde{\mathbf{R}}_{i}, 
\tilde{X}_{i}, t_{2}\right)  \right], 
\label{34}
\end{eqnarray}
\begin{eqnarray}
&\displaystyle 
\tilde{U}^{(1)}_{iy}\left(\tilde{\mathbf{r}}_{i}, \tilde{X}_{i}, t\right) 
\nonumber 
\\ &\displaystyle = \frac{e_{i}}{m_{i}}\int\limits ^{t}_{t_{0}}dt_{1}\left[-\frac{\partial }
{\partial \tilde{X}_{i}}\left(\tilde{E}_{x}
\left(\tilde{\mathbf{r}}_{i}+\tilde{\mathbf{R}}_{i}, \tilde{X}_{i}, t_{1}\right)\right)
\sin \omega_{ci}\left(t-t_{1}\right)\int\limits^{t_{1}}_{t_{0}}dt_{2}\tilde{U}^{(0)}_{ix}
\left(\tilde{\mathbf{r}}_{i}+\tilde{\mathbf{R}}_{i}, \tilde{X}_{i}, t_{2}\right)\right. 
\nonumber 
\\ &\displaystyle
\left.+ \frac{\partial}{\partial \tilde{X}_{i}}\left(\tilde{E}_{y}
\left(\tilde{\mathbf{r}}_{i}+\tilde{\mathbf{R}}_{i}, \tilde{X}_{i}, t_{1}\right)\right)
\cos \omega_{ci}\left(t-t_{1}\right)\int\limits^{t_{1}}_{t_{0}}dt_{2}
\tilde{U}^{(0)}_{ix}\left(\tilde{\mathbf{r}}_{i}+\tilde{\mathbf{R}}_{i}, 
\tilde{X}_{i}, t_{2}\right)  \right]. 
\label{35}
\end{eqnarray}
With variables $\tilde{\mathbf{v}}_{i}$, $\tilde{X}_{i}, \tilde{Y}_{i}$, $\tilde{\mathbf{r}}_{i}$ 
the Vlasov equation for the distribution function \\ 
$F_{i}\left(\tilde{\mathbf{v}}_{i}, \tilde{X}_{i}, \tilde{Y}_{i}, \tilde{\mathbf{r}}_{i}, t\right)$ 
of ions in the sheared poloidal flow for time $t\gg \tau_{corr} \sim \gamma^{-1}$  becomes
\begin{eqnarray}
&\displaystyle 
\frac{\partial F_{i}}{\partial t} + \tilde{v}_{ix}\frac{\partial F_{i}}{\partial \tilde{X}_{i}} 
+\left(\tilde{v}_{iy}-V'_{0}t\tilde{v}_{ix}\right)\frac{\partial F_{i}}{\partial \tilde{Y}_{i}}- 
\tilde{v}_{ix}\int\limits_{t_{0}}^{t}\frac{\partial}
{\partial X_{i}}\tilde{V}_{ix}\left(\mathbf{r}_{i}, X_{i}, t_{1} \right)dt_{1}
\frac{\partial F_{i}}{\partial \tilde{X}_{i}} 
\nonumber 
\\ &\displaystyle
-\tilde{v}_{ix}\int\limits_{t_{0}}^{t}\frac{\partial}{\partial X_{i}}\tilde{V}_{iy}
\left(\mathbf{r}_{i}, X_{i}, t_{1} \right)dt_{1}
\frac{\partial F_{i}}{\partial \tilde{Y}_{i}} - \tilde{U}_{ix}\left(\tilde{\mathbf{r}}_{i}, 
\tilde{X}_{i}, t\right)\int\limits_{t_{0}}^{t}\frac{\partial}
{\partial X_{i}}\tilde{V}_{ix}\left(\mathbf{r}_{i}, X_{i}, t_{1} \right)dt_{1}
\frac{\partial F_{i}}{\partial \tilde{X}_{i}}
\nonumber 
\\ &\displaystyle 
-\tilde{U}_{ix}\left(\tilde{\mathbf{r}}_{i}, \tilde{X}_{i}, t\right)\left(V'_{0}t
+\int\limits_{t_{0}}^{t}\frac{\partial}{\partial X_{i}}\tilde{V}_{iy}
\left(\mathbf{r}_{i}, X_{i}, t_{1} \right)dt_{1}\right)\frac{\partial F_{i}}
{\partial\tilde{Y}_{i}}
\nonumber 
\\ &\displaystyle
+\omega_{ci}\tilde{v}_{iy}\frac{\partial F_{i}}{\partial \tilde{v}_{ix}} -\left(\omega_{ci}
+V'_{0}\right)\tilde{v}_{ix}\frac{\partial F_{i}}{\partial \tilde{v}_{iy}}
\nonumber 
\\ &\displaystyle
-\frac{e_{i}}{m_{i}}\left(\frac{\partial}{\partial \tilde{X}_{i}}
\varphi_{i}\left(\tilde{\mathbf{r}}_{i}+\tilde{\mathbf{R}}_{i}, \tilde{X}_{i}, 
\tilde{Y}_{i}, t \right)
- V'_{0}t \frac{\partial }{\partial \tilde{Y}_{i}}\varphi_{i}\left(\tilde{\mathbf{r}}_{i}+
\tilde{\mathbf{R}}_{i}, \tilde{X}_{i}, \tilde{Y}_{i}, t \right)\right)\frac{\partial F_{i}}{\partial 
\tilde{v}_{ix}}
\nonumber 
\\ &\displaystyle
-\frac{e_{i}}{m_{i}}\frac{\partial }{\partial \tilde{Y}_{i}}\varphi_{i}\left(\tilde{\mathbf{r}}_{i}+
\tilde{\mathbf{R}}_{i}, \tilde{X}_{i}, \tilde{Y}_{i}, t \right)
\frac{\partial F_{i}}{\partial \tilde{v}_{iy}}
-\frac{e_{i}}{m_{i}}\frac{\partial }{\partial \tilde{z}}\varphi_{i}\left(\tilde{\mathbf{r}}_{i}+
\tilde{\mathbf{R}}_{i}, \tilde{X}_{i}, \tilde{Y}_{i}, t\right)
\frac{\partial F_{i}}{\partial v_{iz}}=0.
\label{36}
\end{eqnarray}
The electrostatic potential $\varphi_{i}$ in Eq. (\ref{36}) depends on the micro- and mesoscales 
and can be expressed in the form
\begin{eqnarray}
&\displaystyle
\varphi_{i}\left(\tilde{\mathbf{r}}_{i}+\tilde{\mathbf{R}}_{i}, \tilde{X}_{i}, \tilde{Y}_{i}, t
\right)= \tilde{\varphi}_{i}
\left(\tilde{\mathbf{r}}_{i}+\tilde{\mathbf{R}}_{i}, \tilde{X}_{i}, t\right)+\Phi_{i}\left(\tilde{X}
_{i}, \tilde{Y}_{i}, t\right),
\label{37}
\end{eqnarray}
where $\tilde{\varphi}_{i}$ is the potential of the microscale turbulence, and $\Phi_{i}
\left(\tilde{X}_{i}, \tilde{Y}_{i}, t\right)$ is the potential of the electrostatic mesoscale 
perturbations. The potential $\varphi\left(\tilde{\mathbf{r}}_{i}+\tilde{\mathbf{R}}_{i}, 
\tilde{X}_{i}, \tilde{Y}_{i}, t\right)$, averaged over the initial phases of the microscale 
perturbations, is 
\begin{eqnarray}
&\displaystyle 
\langle\varphi_{i}\left(\tilde{\mathbf{r}}_{i}+\tilde{\mathbf{R}}_{i}, \tilde{X}_{i}, 
\tilde{Y}_{i}, t\right)\rangle=
\Phi_{i}\left(\tilde{X}_{i}, \tilde{Y}_{i}, t\right),
\label{38}
\end{eqnarray}
because $\langle\tilde{\varphi}_{i}\left(\tilde{\mathbf{r}}, \tilde{X}_{i}, t\right)\rangle=0$. 
It follows from Eq. (\ref{27}), that the equation for the Eulerian mean velocity $\left\langle 
\tilde{\mathbf{V}}_{i}\left(\mathbf{r}_{i}, X_{i}, t\right)\right\rangle$ 
averaged over the initial phases $\theta\left(\mathbf{k}\right)$ , 
\begin{eqnarray}
&\displaystyle 
\frac{\partial\left\langle \tilde{\mathbf{V}}_{i}\left(\mathbf{r}_{i}, X_{i}, t\right)
\right\rangle}{\partial t}
-\frac{e_{i}}{m_{i}c}\left[\left\langle\tilde{\mathbf{V}}_{i}\left(\mathbf{r}_{i}, 
X_{i}, t\right)\right\rangle \times\mathbf{B}_{0}\right]
\nonumber\\ 
&\displaystyle
=-\left\langle \tilde{V}_{ix}\frac{\partial \tilde{\mathbf{V}}_{i}}
{\partial X_{i}}\right\rangle +\frac{e_{i}}{m_{i}}\left\langle \tilde{\mathbf{E}}_{i}
\left(\mathbf{r}_{i}, X_{i}, t \right)\right\rangle,
\label{39}
\end{eqnarray}
has nonlinear "extra term" originated from the convective part of the material time derivative. 
This term is absent in the averaged hybrid Eulerian - Lagrangian  Eq. (\ref{28}) for 
$\mathbf{U}_{i}\left(\tilde{\mathbf{r}}_{i}, \tilde{X}_{i}, t\right)$,
\begin{eqnarray}
&\displaystyle 
\frac{d}{dt}\left\langle \tilde{\mathbf{U}}_{i}\left(\tilde{\mathbf{r}}_{i}, \tilde{X}_{i}, t\right)
\right\rangle-\frac{e_{i}}{m_{i}c}\left[\left\langle \tilde{\mathbf{U}}_{i}\left(\tilde{\mathbf{r}}
_{i}, \tilde{X}_{i}, t\right)\right\rangle \times\mathbf{B}_{0}\right]
\nonumber\\ 
&\displaystyle
=\frac{e_{i}}{m_{i}}\left\langle \tilde{\mathbf{E}}_{i}\left(\tilde{\mathbf{r}}_{i}
+\tilde{\mathbf{R}}_{i}, \tilde{X}_{i}+\tilde{R}_{ix}\left(\tilde{\mathbf{r}}_{i}, 
\tilde{X}_{i}, t\right), t\right)\right\rangle, 
\label{40}
\end{eqnarray}
Contrary to the Eulerian equation, where the 
average is determined at definite $X_{i}$ value, in the hybrid of Eulerian - Lagrangian 
description the average at time $t$ is taken with respect to the  positions  
$\tilde{X}_{i}+ \tilde{R}_{ix}\left(\tilde{\mathbf{r}}_{i}, \tilde{X}_{i}, t\right)$ 
displaced by the microturbulence. The Vlasov equation for the ion distribution function 
$\bar{F}_{i}\left(\tilde{\mathbf{v}}_{i}, 
\tilde{X}_{i}, \tilde{Y}_{i}, t\right)$, averaged over the microscale initial 
phases for a time $t\gg \tau _{corr}\sim \gamma^{-1}$, is
\begin{eqnarray}
&\displaystyle 
\frac{\partial \bar{F}_{i}}{\partial t} +\tilde{v}_{ix}\frac{\partial \bar{F}_{i}}
{\partial \tilde{X}_{i}} +\left(\tilde{v}_{iy}-V'_{0}t\tilde{v}_{ix}\right)
\frac{\partial \bar{F}_{i}}{\partial \tilde{Y}_{i}}-\bar{U}_{ix}\left(\tilde{X}_{i}, t\right)
\frac{\partial \bar{F}_{i}}{\partial \tilde{X}_{i}} -\bar{U}_{iy}\left(\tilde{X}_{i}, t\right)
\frac{\partial \bar{F}_{i}}{\partial \tilde{Y}_{i}}
\nonumber 
\\ &\displaystyle
+\omega_{ci}\tilde{v}_{iy}\frac{\partial \bar{F}_{i}}{\partial \tilde{v}_{ix}}-\left(\omega_{ci}
+V'_{0}\right)\frac{\partial \bar{F}_{i}}{\partial \tilde{v}_{iy}}
\nonumber 
\\ &\displaystyle
-\frac{e_{i}}{m_{i}}\left(\frac{\partial \Phi_{i} \left(\tilde{X}_{i}, \tilde{Y}_{i}, t\right)}
{\partial \tilde{X}_{i}}- V'_{0}t \frac{\partial \Phi_{i} \left(\tilde{X}_{i}, 
\tilde{Y}_{i}, t\right)}{\partial \tilde{Y}_{i}}\right)\frac{\partial F_{i}}{\partial 
\tilde{v}_{ix}}
\nonumber 
\\ &\displaystyle
-\frac{e_{i}}{m_{i}}\frac{\partial \Phi_{i} \left(\tilde{X}_{i}, \tilde{Y}_{i}, t\right)}
{\partial \tilde{Y}_{i} }\frac{\partial F_{i}}{\partial \tilde{v}_{iy}}
-\frac{e_{i}}{m_{i}}\frac{\partial \Phi_{i} \left(\tilde{X}_{i}, \tilde{Y}_{i}, t\right)}
{\partial \tilde{z}}\frac{\partial F_{i}}{\partial v_{iz}}=0.
\label{41}
\end{eqnarray}
In this equation, we use the relations 
\begin{eqnarray}
&\displaystyle 
\frac{\partial}{\partial X_{i}}\tilde{V}_{ix}\left(\mathbf{r}_{i}, X_{i}, 
t_{1} \right)=\frac{\partial}{\partial X_{i}}\tilde{U}_{ix}
\left(\tilde{\mathbf{r}}_{i}, \tilde{X}_{i}, t_{1} \right )
\nonumber 
\\ &\displaystyle
\approx \frac{\partial}{\partial\tilde{X}_{i}}\left(\tilde{U}_{ix}\left(\tilde{\mathbf{r}}_{i}, 
\tilde{X}_{i}, t_{1} \right)\right)
\left(1-\int\limits_{t_{0}}^{t}\frac{\partial}{\partial\tilde{X}_{i}}
\tilde{U}_{ix}\left(\tilde{\mathbf{r}}_{i}, \tilde{X}_{i}, t_{1} \right)dt_{1}\right)
\label{42}
\end{eqnarray}
and
\begin{eqnarray}
&\displaystyle 
\frac{\partial}{\partial X_{i}}\tilde{V}_{iy}\left(\mathbf{r}_{i}, X_{i}, 
t_{1} \right)=\frac{\partial}{\partial X_{i}}\tilde{U}_{iy}
\left(\tilde{\mathbf{r}}_{i}, \tilde{X}_{i}, t_{1} \right )
\nonumber 
\\ &\displaystyle
\approx \frac{\partial}{\partial\tilde{X}_{i}}\left(\tilde{U}_{iy}\left(\tilde{\mathbf{r}}_{i}, 
\tilde{X}_{i}, t_{1} \right)\right)
\left(1-\int\limits_{t_{0}}^{t}\frac{\partial}{\partial\tilde{X}_{i}}
\tilde{U}_{ix}\left(\tilde{\mathbf{r}}_{i}, \tilde{X}_{i}, t_{1} \right)dt_{1}\right).
\label{43}
\end{eqnarray}
These relations stem from the identity $\tilde{\mathbf{V}}_{i}\left(\mathbf{r}_{i}, 
X_{i}, t\right)=\tilde{\mathbf{U}}_{i}
\left(\tilde{\mathbf{r}}_{i}, \tilde{X}_{i}, t\right)$, which follows from Eqs. (\ref{19}) and 
(\ref{23}), and from the identity 
\begin{eqnarray*}
&\displaystyle 
\frac{\partial \tilde{X}_{i}}{\partial X_{i}}=1-\int \limits^{t}_{t_{0}}dt_{1}
\frac{\partial \tilde{V}_{ix}\left(\mathbf{r}_{i}, 
X_{i}, t_{1}\right)}{\partial X_{i}},
\end{eqnarray*}
which follows from Eqs. (\ref{20}) and (\ref{24}). In Eq. (\ref{41}), 
$\bar{U}_{ix}\left(\tilde{X}_{i}, t\right)$ and $\bar{U}_{iy}
\left(\tilde{X}_{i}, t\right)$ are the spatially inhomogeneous velocities of the 
convective flows along directions of $x_{i}$ and $y_{i}$, respectively. The velocity $\bar{U}_{ix}
\left(\tilde{X}_{i}, t\right)$ with accounting for the terms on the order of 
$O\left(\left|\tilde{R}_{ix}/L_{\tilde{E}}\right|^{3}\right)$ is determined by the equation
\begin{eqnarray}
&\displaystyle 
\bar{U}_{ix}\left(\tilde{X}_{i}, t\right)=\left\langle \tilde{U}_{ix}\left(\tilde{\mathbf{r}}_{i}, 
\tilde{X}_{i}, t\right)\int\limits_{t_{0}}^{t}\frac{\partial}
{\partial X_{i}}\tilde{V}_{ix}\left(\mathbf{r}_{i}, X_{i}, t_{1} \right)
dt_{1} \right\rangle 
\nonumber 
\\ &\displaystyle
\approx \bar{U}^{(0)}_{ix}\left(\tilde{X}_{i}, t\right)+\bar{U}^{(1)}_{ix}
\left(\tilde{X}_{i}, t\right)+\bar{U}^{(2)}_{ix}\left(\tilde{X}_{i}, t\right),
\label{44}
\end{eqnarray}
where
\begin{eqnarray}
&\displaystyle
\bar{U}^{(0)}_{ix}\left(\tilde{X}_{i}, t\right)=\left\langle \tilde{U}^{(0)}_{ix}
\left(\tilde{\mathbf{r}}_{i}, \tilde{X}
_{i}, t\right)\int\limits_{t_{0}}^{t}\frac{\partial}
{\partial\tilde{X}_{i}}\tilde{U}^{(0)}_{ix}\left(\tilde{\mathbf{r}}_{i}, \tilde{X}_{i}, t_{1} \right)
dt_{1} \right\rangle, 
\label{45}
\end{eqnarray}
\begin{eqnarray}
&\displaystyle
\bar{U}^{(1)}_{ix}\left(\tilde{X}_{i}, t\right)=\left\langle \tilde{U}^{(0)}_{ix}
\left(\tilde{\mathbf{r}}_{i}, \tilde{X}
_{i}, t\right)\int\limits_{t_{0}}^{t}\frac{\partial}
{\partial\tilde{X}_{i}}\tilde{U}^{(1)}_{ix}\left(\tilde{\mathbf{r}}_{i}, 
\tilde{X}_{i}, t_{1} \right)dt_{1} \right\rangle
\nonumber 
\\ &\displaystyle
+\left\langle \tilde{U}^{(1)}_{ix}
\left(\tilde{\mathbf{r}}_{i}, \tilde{X}
_{i}, t\right)\int\limits_{t_{0}}^{t}\frac{\partial}
{\partial\tilde{X}_{i}}\tilde{U}^{(0)}_{ix}\left(\tilde{\mathbf{r}}_{i}, 
\tilde{X}_{i}, t_{1} \right)dt_{1} \right\rangle,
\label{46}
\end{eqnarray}
\begin{eqnarray}
&\displaystyle
\bar{U}^{(2)}_{ix}\left(\tilde{X}_{i}, t\right)=-\left\langle \tilde{U}^{(0)}_{ix}
\left(\mathbf{\tilde{r}}_{i}, \tilde{X}_{i}, t\right)\right.
\nonumber 
\\ &\displaystyle
\left.\times\int\limits_{t_{0}}^{t}\frac{\partial}
{\partial\tilde{X}_{i}}\tilde{U}^{(0)}_{ix}\left(\tilde{\mathbf{r}}_{i}, 
\tilde{X}_{i}, t_{1} \right)dt_{1} \int\limits_{t_{0}}^{t_{1}}\frac{\partial}
{\partial\tilde{X}_{i}}\tilde{U}^{(0)}_{ix}\left(\tilde{\mathbf{r}}_{i}, 
\tilde{X}_{i}, t_{2} \right)dt_{2} \right\rangle.
\label{47}
\end{eqnarray}
Velocity $\bar{U}_{iy}\left(\tilde{X}_{i}, t\right)$ is expressed as
\begin{eqnarray}
&\displaystyle 
\bar{U}_{iy}\left(\tilde{X}_{i}, t\right)=\left\langle \tilde{U}_{ix}\left(\tilde{\mathbf{r}}_{i}, 
\tilde{X}_{i}, t\right)\int\limits_{t_{0}}^{t}\frac{\partial}
{\partial X_{i}}\tilde{V}_{iy}\left(\mathbf{r}_{i}, X_{i}, t_{1} \right)dt_{1}\right\rangle 
\nonumber 
\\ &\displaystyle
\approx \bar{U}^{(0)}_{iy}\left(\tilde{X}_{i}, t\right)+\bar{U}^{(1)}_{iy}
\left(\tilde{X}_{i}, t\right)+\bar{U}^{(2)}_{iy}\left(\tilde{X}_{i}, t\right),
\label{48}
\end{eqnarray}
where
\begin{eqnarray}
&\displaystyle
\bar{U}^{(0)}_{iy}\left(\tilde{X}_{i}, t\right)=\left\langle \tilde{U}^{(0)}_{ix}
\left(\tilde{\mathbf{r}}_{i}, \tilde{X}
_{i}, t\right)\int\limits_{t_{0}}^{t}\frac{\partial}
{\partial\tilde{X}_{i}}\tilde{U}^{(0)}_{iy}\left(\tilde{\mathbf{r}}_{i}, \tilde{X}_{i}, t_{1} \right)
dt_{1} \right\rangle, 
\label{49}
\end{eqnarray}
\begin{eqnarray}
&\displaystyle
\bar{U}^{(1)}_{iy}\left(\tilde{X}_{i}, t\right)=\left\langle \tilde{U}^{(0)}_{ix}
\left(\tilde{\mathbf{r}}_{i}, \tilde{X}
_{i}, t\right)\int\limits_{t_{0}}^{t}\frac{\partial}
{\partial\tilde{X}_{i}}\tilde{U}^{(1)}_{iy}\left(\tilde{\mathbf{r}}_{i}, 
\tilde{X}_{i}, t_{1} \right)dt_{1} \right\rangle
\nonumber 
\\ &\displaystyle
+\left\langle \tilde{U}^{(1)}_{ix}
\left(\tilde{\mathbf{r}}_{i}, \tilde{X}
_{i}, t\right)\int\limits_{t_{0}}^{t}\frac{\partial}
{\partial\tilde{X}_{i}}\tilde{U}^{(0)}_{iy}\left(\tilde{\mathbf{r}}_{i}, 
\tilde{X}_{i}, t_{1} \right)dt_{1} \right\rangle,
\label{50}
\end{eqnarray}
\begin{eqnarray}
&\displaystyle
\bar{U}^{(2)}_{iy}\left(\tilde{X}_{i}, t\right)=-\left\langle \tilde{U}^{(0)}_{ix}
\left(\tilde{\mathbf{r}}_{i}, \tilde{X}_{i}, t\right)\right.
\nonumber 
\\ &\displaystyle
\left.\times\int\limits_{t_{0}}^{t}\frac{\partial}
{\partial\tilde{X}_{i}}\tilde{U}^{(0)}_{iy}\left(\tilde{\mathbf{r}}_{i}, 
\tilde{X}_{i}, t_{1} \right)dt_{1} \int\limits_{t_{0}}^{t_{1}}\frac{\partial}
{\partial\tilde{X}_{i}}\tilde{U}^{(0)}_{ix}\left(\tilde{\mathbf{r}}_{i}, 
\tilde{X}_{i}, t_{2} \right)dt_{2} \right\rangle.
\label{51}
\end{eqnarray}
\end{widetext}
For the electric field $\tilde{\mathbf{E}}_{i}$, given by Eq. (\ref{16}), 
the velocities $\bar{U}^{(0)}_{ix}\left(\tilde{X}_{i}, t\right)$ and $\bar{U}^{(0)}_{iy}
\left(\tilde{X}_{i}, t\right)$ are determined in the Appendix. 

The Vlasov equation for the average electron distribution $\bar{F}_{e}
\left(\tilde{\mathbf{v}}_{e}, \tilde{X}_{e}, \tilde{Y}_{e}, t\right)$, where 
$\tilde{X}_{e}, \tilde{Y}_{e}$ are determined by Eqs. (\ref{19}) - (\ref{25}) with ion species 
subscript changed on electron, has a form similar to 
Eq. (\ref{41}). The velocities $\bar{U}_{ex}\left(\tilde{X}_{e}, t\right)$ and $\bar{U}_{ey}
\left(\tilde{X}_{e}, t\right)$ are determined in this equation by Eqs. (\ref{44}) - (\ref{53}) in 
which velocities $\tilde{U}_{ex}\left(\tilde{\mathbf{r}}_{e}, \tilde{X}_{e}, t\right)$ and 
$\tilde{U}_{ey}\left(\tilde{\mathbf{r}}_{e}, \tilde{X}_{e}, t\right)$ are determined by Eqs. 
(\ref{30}) - (\ref{35}) with the turbulent electric field $\tilde{\mathbf{E}}_{e}
\left(\tilde{\mathbf{r}}_{e}, \tilde{X}_{e}, t\right)$, determined by Eq. (\ref{17}) .

The Vlasov equations (\ref{41}) for $\bar{F}_{i}
\left(\tilde{\mathbf{v}}_{i}, \tilde{X}_{i}, \tilde{Y}_{i}, t\right)$, 
and the similar equation for \\ $\bar{F}_{e}
\left(\tilde{\mathbf{v}}_{e}, \tilde{X}_{e}, \tilde{Y}_{e}, t\right)$, 
and the Poisson equation for the potential $\Phi_{i}\left(\tilde{X}_{i}, \tilde{Y}_{i}, 
t\right)$,
\begin{eqnarray}
&\displaystyle 
\frac{\partial^{2} \Phi_{i}\left(\tilde{X}_{i}, \tilde{Y}_{i}, t\right)}
{\partial^{2} \tilde{X_{i}}}+ \frac{\partial^{2}
 \Phi_{i}\left(\tilde{X}_{i}, \tilde{Y}_{i}, t\right)}
{\partial^{2}\tilde{Y_{i}}}=-4\pi\left(e_{i}\int d\mathbf{v}_{i}\bar{F}_{i}
\left(\tilde{\mathbf{v}}_{i}, \tilde{X}_{i}, \tilde{Y}_{i}, t\right)\right.
\nonumber 
\\ &\displaystyle
\left.-|e|\int d\mathbf{v}_{e}\bar{F}_{e}
\left(\tilde{\mathbf{v}}_{e}, \tilde{X}_{e}, \tilde{Y}_{e}, t\right)\right),
\label{52}
\end{eqnarray}
compose the Vlasov-Poisson system, which governs the kinetic  mesoscale evolution of a plasma 
under the average action of the spatially inhomogeneous microturbulence. 

\section{Hydrodynamics of the mesoscale convective flows}\label{sec4} 
The mesoscale spatial variations of a plasma are involved in the theory of the 
microturbulence through the dependences on $X_{i}$ and on $X_{e}$ of the ion 
and electron densities and temperatures. 
In this Section, we derive a closed set of equations which
determine the mesoscale evolution of the ion and electron densities, velocities, 
temperatures and of the electrostatic potential in the poloidal sheared flow with 
radially inhomogeneous convective flows. 
These equations are the first three moments of the Vlasov equations (\ref{41}) 
for $\bar{F}_{i}$ and $\bar{F}_{e}$. Starting with Eq. (\ref{41}) we 
derive the moment equations as follows. By integrating over velocities 
$\tilde{\mathbf{v}}_{i}$ we derive the ion particles density conservation equation:
\begin{eqnarray}
&\displaystyle 
\frac{\partial n_{i}}{\partial t} +\frac{\partial}{\partial \tilde{X}_{i}}\left(n_{i}u_{ix}\right)
+\frac{\partial}{\partial \tilde{Y}_{i}}\left(n_{i}\left(u_{iy}-V'_{0}tu_{ix}\right)\right)
\nonumber 
\\ &\displaystyle
- \bar{U}_{ix}\left(\tilde{X}_{i}, t\right)\frac{\partial n_{i}}{\partial \tilde{X}_{i}}-\bar{U}_{iy}
\left(\tilde{X}_{i}, t\right)
\frac{\partial n_{i}}{\partial\tilde{Y}_{i}}=0.
\label{53}
\end{eqnarray}
By first multiplying by $\tilde{\mathbf{v}}_{i}$ of Eq. (\ref{41}) and integrating over velocities 
$\tilde{\mathbf{v}}_{i}$, we obtain the momentum conservation equations:
\begin{widetext}
\begin{eqnarray}
&\displaystyle 
\frac{\partial u_{ix}}{\partial t}+ u_{ix}\frac{\partial u_{ix}}{\partial \tilde{X}_{i}}
+\left(u_{iy}-V'_{0}tu_{ix}\right)\frac{\partial u_{ix}}{\partial 
\tilde{Y}_{i}}
-\bar{U}_{ix}\left(\tilde{X}_{i}, t\right)\frac{\partial u_{ix}}{\partial \tilde{X}_{i}}
-\bar{U}_{iy}\left(\tilde{X}_{i}, t\right)\frac{\partial u_{ix}}
{\partial \tilde{Y}_{i}}
\nonumber 
\\ &\displaystyle
=-\frac{1}{m_{i}n_{i}}\left(\frac{\partial P_{i}}{\partial \tilde{X}_{i}}
-V'_{0}t\frac{\partial P_{i}}{\partial \tilde{Y}_{i}}\right)
-\frac{e_{i}}{m_{i}}\left(\frac{\partial \Phi\left(\tilde{X}_{i}, \tilde{Y}_{i}, t\right)}
{\partial \tilde{X}_{i}}-
V'_{0}t\frac{\partial \Phi\left(\tilde{X}_{i}, \tilde{Y}_{i}, t\right)}
{\partial \tilde{Y}_{i}}\right)+\omega_{ci}u_{iy},
\label{54}
\end{eqnarray}
\begin{eqnarray}
&\displaystyle 
\frac{\partial u_{iy}}{\partial t}+ u_{ix}\frac{\partial u_{iy}}{\partial \tilde{X}_{i}}
+\left(u_{iy}-V'_{0}tu_{ix}\right)\frac{\partial u_{iy}}{\partial 
\tilde{Y}_{i}}-\bar{U}_{ix}\left(\tilde{X}_{i}, t\right)\frac{\partial u_{iy}}
{\partial \tilde{X}_{i}} -\bar{U}_{iy}
\left(\tilde{X}_{i}, t\right)\frac{\partial u_{iy}}{\partial \tilde{Y}_{i}}
\nonumber 
\\ &\displaystyle
=-\frac{1}{m_{i}n_{i}}\frac{\partial P_{i}}
{\partial\tilde{Y}_{i}}
-\frac{e_{i}}{m_{i}}\frac{\partial \Phi\left(\tilde{X}_{i}, \tilde{Y}_{i}, t\right)}
{\partial \tilde{Y}_{i}}-\left(\omega_{ci}+V'_{0}\right)u_{ix}.
\label{55}
\end{eqnarray}
By first multiplying Eq. (\ref{41}) by $\frac{1}{2}\left|\tilde{\mathbf{v}}_{i}
-\mathbf{u}_{i}\left(\tilde{X}_{i}, \tilde{Y}_{i}, t\right)\right|^{2}$ 
and integrating over velocities $\tilde{\mathbf{v}}_{i}$ we obtain the equation 
\begin{eqnarray}
&\displaystyle 
\frac{\partial T_{i}}{\partial t}+u_{ix}\frac{\partial T_{i}}
{\partial \tilde{X}_{i}}+\left(u_{iy}-V'_{0}tu_{ix}\right)\frac{\partial T_{i}}
{\partial \tilde{Y}_{i}}-\bar{U}_{ix}\left(\tilde{X}_{i}, t\right)\frac{\partial T_{i}}
{\partial\tilde{X}_{i}} -\bar{U}_{iy}\left(\tilde{X}_{i}, t\right)
\frac{\partial T_{i}}{\partial\tilde{Y}_{i}}
\nonumber 
\\ &\displaystyle
= -\frac{2}{3}T_{i}\left(\frac{\partial u_{ix}}{\partial \tilde{X}_{i}}-V'_{0}t
\frac{\partial u_{ix}}{\partial \tilde{Y}_{i}}+\frac{\partial u_{iy}}{\partial 
\tilde{Y}_{i}}\right),
\label{56}
\end{eqnarray}
\end{widetext}
which determines the evolution of the ion temperature in the poloidal sheared flow with 
convective flows. The definitions of the ion number density $n_{i}
\left(\tilde{X}_{i}, \tilde{Y}_{i}, t\right)$,
the velocity $\mathbf{u}_{i}\left(\tilde{X}_{i}, \tilde{Y}_{i}, t\right)$ 
and the ion temperature $T_{i}\left(\tilde{X}_{i}, \tilde{Y}_{i}, t\right)$ 
are the usual ones:
\begin{eqnarray}
&\displaystyle 
n_{i}\left(\tilde{X}_{i}, \tilde{Y}_{i}, t\right)=\int d\tilde{\mathbf{v}}_{i}\bar{F}_{i}
\left(\tilde{\mathbf{v}}_{i}, \tilde{X}_{i}, 
\tilde{Y}_{i}, t\right),
\label{57}
\end{eqnarray}
\begin{eqnarray}
&\displaystyle 
\mathbf{u}_{i}\left(\tilde{X}_{i}, \tilde{Y}_{i}, t\right)
=\frac{1}{n_{i}\left(\tilde{X}_{i}, 
\tilde{Y}_{i},t\right)}\int d\tilde{\mathbf{v}}_{i}
\tilde{\mathbf{v}}_{i}\bar{F}_{i},
\label{58}
\end{eqnarray}
\begin{eqnarray}
&\displaystyle 
T_{i}\left(\tilde{X}_{i}, \tilde{Y}_{i}, t\right)=\frac{m_{i}}{3n_{i}\left(\tilde{X}_{i}, 
\tilde{Y}_{i}, t\right)}\int d\tilde{\mathbf{v}}_{i}
\left|\tilde{\mathbf{v}}_{i}-\mathbf{u}_{i}\right|^{2}\bar{F}_{i},
\label{59}
\end{eqnarray}
and $P_{i}=n_{i}T_{i}$ is the ion thermal pressure. Equations (\ref{50}) - (\ref{53}) for 
ions and the similar equations for electrons, and the 
Poisson equation for the potential $\Phi_{i}\left(\tilde{X}_{i}, \tilde{Y}_{i}, t\right)$,
\begin{eqnarray}
&\displaystyle 
\frac{\partial^{2} \Phi_{i}}{\partial^{2} \tilde{X_{i}}}+ \frac{\partial^{2} \Phi_{i}}
{\partial^{2}\tilde{Y_{i}}}=
-4\pi\left(e_{i}n_{i}\left(\tilde{X}_{i}, \tilde{Y}_{i}, t\right)
-|e|n^{(i)}_{e}\left(\tilde{X}_{i}, \tilde{Y}_{i}, t\right)\right),
\label{60}
\end{eqnarray}
where $n^{(i)}_{e}\left(\tilde{X}_{i}, \tilde{Y}_{i}, t\right)$ is the electron density 
perturbation, determined in the ion frame variables $\tilde{X}_{i}, \tilde{Y}_{i}$, 
in which potential $\Phi_{i}\left(\tilde{X}_{i}, \tilde{Y}_{i}, t\right)$ in Eq. (\ref{60}) 
is determined, compose the basic system of the fluid equations, which 
governs the mesoscale evolution of the poloidal sheared plasma flow with radially 
inhomogeneous convective flows across the magnetic field.

As it follows from Eqs. (\ref{95}), (\ref{96}) and (\ref{99}), (\ref{100}), the velocities 
$\bar{\mathbf{U}}_{i}\left(\tilde{X}_{i}, t\right)$, $\bar{\mathbf{U}}_{e}
\left(\tilde{X}_{i}, t\right)$ of the ion and electron convective flows are 
proportional to the mesoscale gradient of the spectral intensity $\left|\varphi_{i}
\left(\mathbf{k}, \tilde{X}_{i}\right)\right|^{2}$ of the turbulent electric field. 
Therefore the maximum of the ion convective flow velocities could be in the pedestal, 
where the maximum gradients of the plasma parameters 
are observed, i. e. for the $\tilde{X}_{i}$ values in the interval $\tilde{X}_{iT} > 
\tilde{X}_{i}> \tilde{X}_{iB}$, where $\tilde{X}_{iT}$ and $\tilde{X}_{iB}$ are the
coordinates of the pedestal top and of the pedestal bottom, respectively. 
Outside this interval, i. e. for the plasma core, 
$\tilde{X}_{i}> \tilde{X}_{iT}$, and in the far SOL, $\tilde{X}_{i}<\tilde{X}_{iB}$, 
the plasma parameters and of FW field are much more uniform and the convective 
flow velocities are more slower. Thus, the problem of the mesoscale evolution of 
the pedestal plasma involves the problem of the stability of a plasma with 
convective flows with spatially inhomogeneous velocities. 
As a first step to the solution of the system of Eqs. (\ref{53}) - (\ref{56}), 
we find the characteristics of the operator 
\begin{eqnarray}
&\displaystyle 
\frac{D}{Dt}=\frac{\partial }{\partial t} -\bar{U}_{ix}\left(\tilde{X}_{i}, t\right)
\frac{\partial }
{\partial \tilde{X}_{i}} -\bar{U}_{iy}\left(\tilde{X}_{i}, t\right)\frac{\partial }
{\partial \tilde{Y}_{i}},
\label{61}
\end{eqnarray}
which are determined by the system
\begin{eqnarray}
&\displaystyle 
dt=-\frac{d\tilde{X}_{i}}{\bar{U}_{ix}\left(\tilde{X}_{i}, t\right)}=-\frac{d\tilde{Y}_{i}}
{\bar{U}_{iy}\left(\tilde{X}_{i}, t\right)}.
\label{62}
\end{eqnarray}
For deriving the simplest solution to system (\ref{62}), which reveals the effects of 
the spatial inhomogeneity of the convective flow velocities, we use in Eq. (\ref{62}) the expansions
\begin{eqnarray}
&\displaystyle \bar{U}_{ix}\left(\tilde{X}_{i}, t\right)=\bar{U}^{(0)}_{ix}+\bar{U}'_{ix}
\left(\tilde{X}^{(0)}_{i}, 
t\right)\left(\tilde{X}_{i}-\tilde{X}^{(0)}_{i}\right), 
\label{63}
\end{eqnarray}
and 
\begin{eqnarray}
&\displaystyle \bar{U}_{iy}\left(\tilde{X}_{i}, t\right) =\bar{U}^{(0)}_{iy}+ \bar{U}'_{iy}
\left(\tilde{X}^{(0)}_{i}, t\right)\left(\tilde{X}_{i}-\tilde{X}^{(0)}_{i}\right)
\label{64}
\end{eqnarray}
at the vicinity of an arbitrary coordinate $\tilde{X}^{(0)}_{i}$, 
and consider the case of the time-independent velocity compressing rate,  
$\bar{U}'_{ix}= const$,  and of the velocity shearing rate, $\bar{U}'_{iy}= const$. 
The solution to system (\ref{62}) for the case of the flows with stationary uniform 
compressing and shearing rates is simple and has a form
\begin{eqnarray}
&\displaystyle 
\check{X}_{i}=\frac{1}{\bar{U}'_{ix}}\left[\left(\bar{U}^{(0)}_{ix}+\bar{U}'_{ix}\left(\tilde{X}_{i}-
\tilde{X}^{(0)}_{i}\right) 
\right)e^{\bar{U}'_{ix}t}- \bar{U}^{(0)}_{ix} \right], 
\label{65}
\end{eqnarray}
and
\begin{eqnarray}
&\displaystyle 
\check{Y}_{i}=\tilde{Y}_{i}+\left(\bar{U}^{(0)}_{iy}-\bar{U}^{(0)}_{ix}\frac{\bar{U}'_{iy}}
{\bar{U}'_{ix}}\right)t
-\frac{\bar{U}'_{iy}}{\left(\bar{U}'_{ix}\right)^{2}}\left(\bar{U}^{(0)}_{ix}
+\bar{U}'_{ix}\left(\tilde{X}_{i}-\tilde{X}^{(0)}_{i}\right)\right),
\label{66}
\end{eqnarray}
$\check{X}_{i}$ and $\check{Y}_{i}$ are the integrals of system (\ref{62}) with 
expansions (\ref{63}), (\ref{64}). Note, that at $t=0$ $\check{X}_{i}=\tilde{X}_{i}-\tilde{X}^{(0)}
_{i}$.

Now we perform the transformations  of variables 
$\tilde{X}_{i}$, $\tilde{Y}_{i}$ in Eqs. (\ref{53}) - (\ref{56}) to variables $\check{X}_{i}$, $
\check{Y}_{i}$ and derive the following equations:
\begin{widetext}
\begin{eqnarray}
&\displaystyle 
\frac{\partial n_{i}\left(\check{X}_{i}, \check{Y}_{i}, t\right)}{\partial t} +e^{\bar{U}'_{ix}t}
\frac{\partial}{\partial \check{X}_{i}}\left(n_{i}u_{ix}
\left(\check{X}_{i}, \check{Y}_{i}, t\right)\right)
\nonumber 
\\ &\displaystyle
-\frac{\bar{U}'_{iy}}{\bar{U}'_{ix}}\frac{\partial}{\partial \check{Y}_{i}}\left(n_{i}u_{ix}
\left(\check{X}_{i}, \check{Y}_{i}, t\right)\right)+
\frac{\partial}{\partial \check{Y}_{i}}\left(n_{i}\left(u_{iy}\left(\check{X}_{i}, \check{Y}_{i}, t
\right)-V'_{0}tu_{ix}\right)\right)=0,
\label{67}
\end{eqnarray}
\begin{eqnarray}
&\displaystyle 
\frac{\partial u_{ix}}{\partial t}+ u_{ix}\left(e^{\bar{U}'_{ix}t}\frac{\partial u_{ix}}
{\partial \check{X}_{i}}-\frac{\bar{U}'_{iy}}{\bar{U}'_{ix}}\frac{\partial u_{ix}}
{\partial \check{Y}_{i}}\right)+\left(u_{iy}-V'_{0}tu_{ix}\right)
\frac{\partial u_{ix}}{\partial \check{Y}_{i}}
\nonumber 
\\ &\displaystyle
=-\frac{1}{m_{i}n_{i}\left(\check{X}_{i}, \check{Y}_{i}, t\right)}\left(e^{\bar{U}'_{ix}t}
\frac{\partial P_{i}}{\partial \check{X}_{i}}
-\frac{\bar{U}'_{iy}}{\bar{U}'_{ix}}\frac{\partial P_{i}}{\partial \check{Y}_{i}}
-V'_{0}t\frac{\partial P_{i}}{\partial \check{Y}_{i}}\right)
\nonumber 
\\ &\displaystyle
-\frac{e_{i}}{m_{i}}\left(e^{\bar{U}'_{ix}t}\frac{\partial }{\partial \check{X}_{i}}
\Phi\left(\check{X}_{i}, \check{Y}_{i}, t\right)
- \frac{\bar{U}'_{iy}}{\bar{U}'_{ix}}\frac{\partial }{\partial \check{Y}_{i}}
\Phi\left(\check{X}_{i}, \check{Y}_{i}, t\right)-
V'_{0}t\frac{\partial }{\partial \check{Y}_{i}}\Phi\left(\check{X}_{i}, \check{Y}_{i}, t\right) 
\right)+\omega_{ci}u_{iy},
\label{68}
\end{eqnarray}

\begin{eqnarray}
&\displaystyle 
\frac{\partial u_{iy}}{\partial t}+ u_{ix}\left(e^{\bar{U}'_{ix}t}\frac{\partial u_{iy}}
{\partial \check{X}
_{i}}-\frac{\bar{U}'_{iy}}{\bar{U}'_{ix}}\frac{\partial u_{iy}}
{\partial \check{Y}_{i}}\right)+\left(u_{iy}-V'_{0}tu_{ix}\right)
\frac{\partial u_{iy}}{\partial \check{Y}_{i}}
\nonumber 
\\ &\displaystyle
=-\frac{1}{m_{i}n_{i}\left(\check{X}_{i}, \check{Y}_{i}, t\right)}\frac{\partial P_{i}}{\partial 
\check{Y}_{i}} -\frac{e_{i}}{m_{i}}\frac{\partial}{\partial \check{Y}_{i}}
\Phi\left(\check{X}_{i}, \check{Y}_{i}, t\right)-\left(\omega_{ci}+V'_{0}\right)u_{ix},
\label{69}
\end{eqnarray}
\begin{eqnarray}
&\displaystyle 
\frac{\partial T_{i}}{\partial t}+u_{ix}\left(e^{\bar{U}'_{ix}t}\frac{\partial T_{i}}{\partial 
\check{X}_{i}}-\frac{\bar{U}'_{iy}}{\bar{U}'_{ix}}\frac{\partial T_{i}}{\partial \check{Y}_{i}}\right)
+\left(u_{iy}-V'_{0}tu_{ix}\right)\frac{\partial T_{i}}{\partial \check{Y}_{i}}
\nonumber 
\\ &\displaystyle
= -\frac{2}{3}T_{i}\left(e^{\bar{U}'_{ix}t}\frac{\partial u_{ix}}{\partial \check{X}_{i}}-
\frac{U'_{iy}}
{\bar{U}'_{ix}}\frac{\partial u_{ix}}{\partial \check{Y}_{i}}+\frac{\partial u_{iy}}
{\partial \check{Y}_{i}}\right),
\label{70}
\end{eqnarray}

\begin{eqnarray}
&\displaystyle
\left(e^{2\bar{U}'_{ix}t}\frac{\partial^{2}}{\partial \check{X}^{2}_{i}}-2e^{\bar{U}'_{ix}t}
\frac{\bar{U}'_{iy}}{\bar{U}'_{ix}}\frac{\partial^{2}}{\partial \check{X}_{i}
\partial \check{Y}_{i}}+ \left(\frac{\bar{U}'_{iy}}{\bar{U}'_{ix}}\right)^{2}
\frac{\partial^{2}}{\partial \check{Y}_{i}^{2}}\right)\Phi\left(\check{X}_{i}, 
\check{Y}_{i}, t\right)
\nonumber 
\\ &\displaystyle
=-4\pi\left(e_{i}n_{i}\left(\check{X}_{i}, \check{Y}_{i}, t\right)
-|e|n^{(i)}_{e}\left(\check{X}_{i}, \check{Y}_{i}, t\right)\right).
\label{71}
\end{eqnarray}
\end{widetext}
This system does not contain any explicit dependences on the radial coordinate $\check{X}_{i}$. 
Instead, these equations involve the explicit time dependences, which reveal the effects 
of the basic poloidal sheared flow and of the convective compressed and sheared flows 
on the mesoscale temporal evolution of the inhomogeneous turbulent 
plasmas. It follows from this system, that the basic poloidal sheared flow with the 
velocity shearing rate $V'_{0}$ leads to the appearance of the linearly growing with time 
coefficients in Eqs. (\ref{67}) - (\ref{71}). This time dependence reveals the effect 
of the continuous distortion with time of the perturbations by the basic poloidal 
sheared flow. This distortion grows with time and forms a time-dependent nonmodal 
processes, the linear and the renormalized nonlinear stages of which were investigated 
in Refs. \cite{Mikhailenko1, Mikhailenko2, Mikhailenko3}.
It was found\cite{Mikhailenko1, Mikhailenko2, Mikhailenko3}, that this nonmodal 
process is at the foundation of the experimentally 
observed effect of the suppression of the drift-type instabilities by the poloidal sheared flow, 
which have the growth rates less than the poloidal flow velocity shearing rate $V'_{0}$. 
Equations (\ref{67}) - (\ref{71}) display also principally different effect of the 
compressed convective flow along $\check{X}_{i}$. It reveals in the exponentially growing 
with time coefficients $e^{\bar{U}'_{ix}t}$ in Eqs. (\ref{67}) - (\ref{71}), which at 
time $t> \left(\bar{U}'_{ix}\right)^{-1}$ determine the compressed flow as
the dominant factor in the evolution of the plasma with a radially inhomogeneous turbulence.
For that time, system (\ref{67}) - (\ref{71}) for $n_{i}, n_{e}$, 
$T_{i}, T_{e}$, $u_{ix}, u_{ex}$, $u_{iy}, u_{ey}$, 
and $\Phi$ can be substantially reduced by neglecting the derivatives over $\check{Y}_{i}$, 
which are exponentially small with respect to the terms containing the derivatives over 
$\check{X}_{i}$. The reduced system, 
\begin{eqnarray}
&\displaystyle 
\frac{\partial n_{i}\left(\check{X}_{i}, t\right)}{\partial t}
+e^{\bar{U}'_{ix}t}\frac{\partial}{\partial \check{X}_{i}}
\left( n_{i}\left(\check{X}_{i}\right)u_{ix}\left(\check{X}_{i}, t\right)\right)=0,
\label{72}
\end{eqnarray}
\begin{eqnarray}
&\displaystyle 
\frac{\partial u_{ix}\left(\check{X}_{i}, t\right)}{\partial t}
+e^{\bar{U}'_{ix}t}u_{ix}\frac{\partial u_{ix}}{\partial \check{X}_{i}}
\nonumber 
\\ &\displaystyle =-\frac{e^{\bar{U}'_{ix}t}}{m_{i}}\left(\frac{1}{n_{i}\left(\check{X}_{i}\right)}
\frac{\partial P_{i}}{\partial \check{X}_{i}}- e_{i}\frac{\partial\Phi\left(\check{X}_{i}, t\right)}
{\partial \check{X}_{i}}\right)+\omega_{ci}u_{iy},
\label{73}
\end{eqnarray}
\begin{eqnarray}
&\displaystyle 
\frac{\partial u_{iy}\left(\check{X}_{i}, t\right)}{\partial t}
+e^{\bar{U}'_{ix}t}u_{ix}\frac{\partial u_{iy}}{\partial \check{X}_{i}}
=-\omega_{ci}u_{ix},
\label{74}
\end{eqnarray}
\begin{eqnarray}
&\displaystyle 
\frac{\partial T_{i}\left(\check{X}_{i}, t\right)}{\partial t}
+e^{\bar{U}'_{ix}t}u_{ix}\frac{\partial T_{i}}{\partial \check{X}_{i}}
=-e^{\bar{U}'_{ix}t}\frac{2}{3}T_{i}
\frac{\partial u_{ix}}{\partial\check{X}_{i}},
\label{75}
\end{eqnarray}
and the  equations for $n_{e}$, $T_{e}$, $u_{ex}$, $u_{ey}$ derived from Eqs. 
(\ref{72}) - (\ref{75}) by changing species index $i$ on $e$, supplemented by the 
equation for the electrostatic potential 
$\Phi\left(\check{X}_{i}, t\right)$,
\begin{eqnarray}
&\displaystyle 
e^{2\bar{U}'_{ix}t}\frac{\partial^{2} \Phi\left(\check{X}_{i}, t\right)}
{\partial \check{X}_{i}^{2} }
=-4\pi\left(e_{i}n_{i}\left(\check{X}_{i}, t\right)
-|e|n^{(i)}_{e}\left(\check{X}_{i}, t\right)\right),
\label{76}
\end{eqnarray}
compose the nonlinear system of equations, which governs the mesoscale dynamics of the 
pedestal plasma. The more simple presentation of all these equations may be derived using
the ion "compressed" time variable $\tau_{i}$, determined by the relation
\begin{eqnarray}
&\displaystyle 
\bar{U}_{ix}'\tau_{i} = e^{\bar{U}_{ix}'t},
\label{77}
\end{eqnarray}
with which Eqs. (\ref{72}) - (\ref{76}) are presented in the following much more simple form:
\begin{eqnarray}
&\displaystyle 
\frac{\partial n_{i}\left(\check{X}_{i}, \tau_{i}\right)}{\partial \tau_{i}}
+\frac{\partial}{\partial \check{X}_{i}}
\left( n_{i}u_{ix}\left(\check{X}_{i}, \tau_{i}\right)\right)=0,
\label{78}
\end{eqnarray}
\begin{eqnarray}
&\displaystyle 
\frac{\partial u_{ix}\left(\check{X}_{i}, \tau_{i}\right)}{\partial \tau_{i}}
+u_{ix}\frac{\partial u_{ix}}{\partial \check{X}_{i}}
\nonumber 
\\ &\displaystyle =-\frac{1}{m_{i}}\left(\frac{1}{n_{i}\left(\check{X}_{i}, \tau_{i}\right)}
\frac{\partial P_{i}}{\partial \check{X}_{i}}
- e_{i}\frac{\partial\Phi\left(\check{X}_{i}, \tau_{i}\right)}{\partial \check{X}_{i}}\right)
+\frac{\omega_{ci}}{\bar{U}_{ix}'\tau}u_{iy}\left(\check{X}_{i}, \tau_{i}\right),
\label{79}
\end{eqnarray}
\begin{eqnarray}
&\displaystyle 
\frac{\partial u_{iy}\left(\check{X}_{i}, \tau_{i}\right)}{\partial \tau_{i}}
+u_{ix}\frac{\partial u_{iy}}{\partial \check{X}_{i}}
=-\frac{\omega_{ci}}{\bar{U}_{ix}'\tau_{i}}u_{ix}\left(\check{X}_{i}, \tau_{i}\right),
\label{80}
\end{eqnarray}
\begin{eqnarray}
&\displaystyle 
\frac{\partial T_{i}\left(\check{X}_{i}, \tau_{i}\right)}{\partial \tau_{i}}
+u_{ix}\frac{\partial T_{i}}{\partial \check{X}_{i}}
+T_{i}\frac{\partial u_{ix}\left(\check{X}_{i}, \tau_{i}\right)}{\partial\check{X}_{i}}=0.
\label{81}
\end{eqnarray}
Using Eqs. (\ref{77}) - (\ref{81}) with species index 
$i$ changed on $e$, we derive the equations for $n_{e}\left(\check{X}_{e}, \tau_{e}\right)$, 
$T_{e}\left(\check{X}_{e}, \tau_{e}\right)$, $u_{ex}\left(\check{X}_{e}, \tau_{e}\right)$, 
$u_{ey}\left(\check{X}_{e}, \tau_{e}\right)$. Also, Eq. (\ref{76}) for the electrostatic potential 
$\Phi\left(\check{X}_{i}, \tau_{i}\right)$, will get a form
\begin{eqnarray}
&\displaystyle 
\frac{\partial^{2} \Phi_{i}\left(\check{X}_{i}, \tau_{i}\right)}
{\partial \check{X}_{i}^{2} }
=\nonumber 
\\ &\displaystyle
=-\frac{4\pi}{\left(\bar{U}_{ix}'\tau_{i}\right)^{2}}\left(e_{i}n_{i}\left(\check{X}_{i}, 
\tau_{i}\right)-|e|n^{(i)}_{e}\left(\check{X}_{i}, \tau_{i}\right)\right).
\label{82}
\end{eqnarray}

For understanding the temporal evolution of the plasma compressing flow it is 
instructive to derive the linearised solutions to system  
(\ref{78}) - (\ref{82}). We present the ion density and the ion 
temperature in Eqs. (\ref{78}) and (\ref{81}) in the forms
\begin{eqnarray}
&\displaystyle 
n_{i}\left(\check{X}_{i}, \tau_{i}\right)=n_{i0}\left(\check{X}_{i}, \tau_{i}\right)
+n_{i1}\left(\check{X}_{i}, \tau_{i}\right),
\label{83}
\end{eqnarray}
and 
\begin{eqnarray}
&\displaystyle 
T_{i}\left(\check{X}_{i}, \tau_{i}\right)=T_{i0}\left(\check{X}_{i}, \tau_{i}\right)
+T_{i1}\left(\check{X}_{i}, \tau_{i}\right),
\label{84}
\end{eqnarray}
where $n_{i0}\left(\check{X}_{i}, \tau_{i}\right)$ and $T_{i0}\left(\check{X}_{i}, 
\tau_{i}\right)$ are the 
equilibrium values of the ion density and of the ion temperature, and $n_{i1}$, $T_{i1}$ 
are their perturbations caused by the self-consistent electrostatic respond of 
a plasma on the relative motion of the ion and electron compressed flows. 
As it follows from Eqs. (\ref{78}) and (\ref{81}), the evolution with time of the ion 
density $n_{i0}\left(\check{X}_{i},\tau_{i}\right)$ 
and of the ion temperature $T_{i0}\left(\check{X}_{i}, \tau_{i}\right)$ in 
compressed flow is determined by the equations   
\begin{eqnarray}
&\displaystyle 
\frac{\partial n_{i0}\left(\check{X}_{i}, \tau_{i}\right)}{\partial \tau_{i}}=0,
\label{85}
\end{eqnarray} 
\begin{eqnarray}
&\displaystyle
\frac{\partial T_{i0}
\left(\check{X}_{i}, \tau_{i}\right)}{\partial \tau_{i}}=0,
\label{86}
\end{eqnarray}
or, for $n_{i0}\left(\tilde{X}_{i}, \tau_{i}\right)$ and  $T_{i0}\left(\tilde{X}_{i}, 
\tau_{i}\right)$,
by the equations 
\begin{eqnarray}
&\displaystyle 
\frac{\partial n_{i0}}{\partial t} - \left(\bar{U}^{(0)}_{ix}+\bar{U}'_{ix}
\left(\tilde{X}_{i}-\tilde{X}^{(0)}_{i}\right)\right)
\frac{\partial n_{i0}}{\partial \tilde{X}_{i}}=0,
\label{87}
\end{eqnarray}
\begin{eqnarray}
&\displaystyle 
\frac{\partial T_{i0}}{\partial t} - \left(\bar{U}^{(0)}_{ix}+\bar{U}'_{ix}
\left(\tilde{X}_{i}-\tilde{X}^{(0)}_{i}\right)\right)
\frac{\partial T_{i0}}{\partial \tilde{X}_{i}}=0,
\label{88}
\end{eqnarray}
where expansion (\ref{63}) was used. The solution to Eq. (\ref{85}),
\begin{eqnarray}
&\displaystyle 
n_{i0}=n_{i0}\left( \check{X}_{i}\right),
\label{89}
\end{eqnarray}
is independent on time, whereas the solution $n_{i0}\left( \tilde{X}_{i}, t\right)$
to Eq. (\ref{87}) for the simplest initial condition 
\begin{eqnarray}
&\displaystyle 
n_{i0}\left(\tilde{X}_{i}, t=0\right)=n_{i0}\left( \check{X}_{i}\right)= 
n_{i0}\left(\tilde{X}_{i0}\right)
+\frac{\partial n_{i0}\left(\tilde{X}_{i}, t=0\right)}{\partial \tilde{X}_{i}}
|_{\tilde{X}_{i}=\tilde{X}_{i0}}\check{X}_{i}
\label{90}
\end{eqnarray}
is equal to 
\begin{eqnarray}
&\displaystyle 
n_{i0}\left(\tilde{X}_{i}, t\right)= n_{i0}\left(\tilde{X}_{i0}\right)
+\frac{\partial n_{i0}\left(\tilde{X}_{i}, t=0\right)} 
{\partial \tilde{X}_{i}}|_{\tilde{X}_{i}=\tilde{X}_{i0}}
\nonumber 
\\ &\displaystyle
\times\frac{1}{\bar{U}'_{ix}}\left[\left(\bar{U}^{(0)}_{ix}+\bar{U}'_{ix}\left(\tilde{X}_{i}
-\tilde{X}^{(0)}_{i}\right)\right)e^{\bar{U}'_{ix}t}- \bar{U}^{(0)}_{ix} \right].
\label{91}
\end{eqnarray}
Equation (\ref{91}) displays that the value of $n_{i0}$ at $\tilde{X}_{i}=\tilde{X}_{i1}$, 
detected at $t=t_{1}$, is shifted by the compressed flow and is
observed in $\tilde{X}_{i} =\tilde{X}_{i2}< \tilde{X}_{i1}$ at time $t=t_{2}>t_{1}$.
The solution similar to Eq. (\ref{91}) is derived also to Eq. (\ref{88}) for the temporal 
evolution of the equilibrium ion temperature $T_{i0}$ in compressed flow. 

In the limit of the vanished compressing rate 
$\bar{U}'_{ix}$, Eq. (\ref{91}) displays the transport of the ion density inhomogeneity 
with velocity $\bar{U}_{ix}\left(\tilde{X}_{i0}\right)$,  
\begin{eqnarray}
&\displaystyle 
n_{i0}\left(\tilde{X}_{i}, t\right)=n_{i0}\left(\tilde{X}_{i0}\right)
+\frac{\partial n_{i0}
\left(\tilde{X}_{i}, t=0\right)}{\partial\tilde{X}_{i}}|_{\tilde{X}_{i}=\tilde{X}_{i0}}\cdot
\left(\tilde{X}_{i} - \tilde{X}_{i0}+\bar{U}^{(0)}_{ix}t\right).
\label{92}
\end{eqnarray}

When the level of IC parametric turbulence is much lover than the level 
of the low frequency drift turbulence, the electric field (\ref{16}) is 
determined by the drift turbulence. The suppression of the edge drift turbulence by 
the poloidal sheared flow entails the formation of the stagnation  point for the 
compressed flow velocity at the pedestal 
bottom, where $\bar{U}_{ix}\left(\tilde{X}_{iB}\right)\approx 0$ and 
$n_{i0}\left(\tilde{X}_{i}<\tilde{X}_{iB}\right) 
\approx 0$. The solution (\ref{91}) for the ion density 
in the region $\tilde{X}_{i}>\tilde{X}_{iB}$ of the pedestal bottom becomes equal to
\begin{eqnarray}
&\displaystyle 
n_{i0}\left(\tilde{X}_{i}, t\right)=
\frac{\partial n_{i0}\left(\tilde{X}_{i}\right)}{\partial \tilde{X}_{i}}|_{\tilde{X}_{i}
=\tilde{X}_{iB}}e^{\bar{U}'_{ix}\left(\tilde{X}_{iB}\right)t}
\left(\tilde{X}_{i}-\tilde{X}_{iB}\right).
\label{93}
\end{eqnarray}
It follows from Eqs. (\ref{93}), that the gradient of the ion density  at 
$\tilde{X}_{i}>\tilde{X}_{iB}$ grows exponentially with time as $e^{\bar{U}'_{ix}t}$. 
This effect of the fast stepping up 
with time of the density profile in the pedestal region by the compressed flow looks like 
the instability development with the growth rate equal to $\bar{U}'_{ix}$ for ions 
and $\bar{U}'_{ex}$ for electrons. 

It follows from Eq. (\ref{73}) that due to the fast growing 
coefficient $e^{\bar{U}'_{ix}t}$ the radial ion pressure force at some time 
$t\gtrsim t_{\star}$ can be larger than the radial component of the ion Lorentz force. 
The crude estimate for the transition time $t_{\star}$ may be derived from the 
balance relation, which follows from Eq. (\ref{73}),
\begin{eqnarray}
&\displaystyle 
e^{\bar{U}'_{ix}t_{\star}}\frac{v_{Ti}}{L_{n}}\sim \omega_{ci},
\label{94}
\end{eqnarray}
where $L_{n}$ is the spatial scale of the ion density gradient of the pedestal plasma.
For the ion temperature $T_{i}\sim 200$ eV, $L_{n}= 2$ cm, and $B_{0}=2$ T, we have 
$e^{\bar{U}'_{ix}t_{\star}}\sim 10$. For the numerical sample, considered in 
Ref. \cite{Mikhailenko}, $\bar{U}_{ix}\sim 0.1v_{Ti}\sim 10^{6}$ cm/s and $\bar{U}'_{ix} 
\sim 5\cdot 10^{5}$ s$^{-1}$ we derive $t_{\star} \sim 6\cdot 10^{-6}$ s. At time $t > t_{\star}$, 
the radial outflow of the temporally unconfined ions forms.

\section{Conclusions}\label{sec5}
In this paper, we develop the kinetic and hydrodynamic theories 
of the convective mesoscale flows, driven by the spatially inhomogeneous IC parametric 
microturbulence of the pedestal plasma with a sheared poloidal flow. The amplitude 
$\tilde{\mathbf{E}}\left(\mathbf{k}, X_{i}\right)$ of 
the electric field (\ref{16}) of the microscale IC 
parametric turbulence of the inhomogeneous plasma, driven by the inhomogeneous FW field, 
is spatially inhomogeneous and depends on the mesoscale coordinate $X_{i}$. 
In this paper, the IC microturbulence is 
assumed to be weakly nonlinear with known dependence of $\tilde{\mathbf{E}}$ on $X_{i}$, 
which established by the linear theory of the IC instability, and by the weak nonlinear theory 
which determines the local saturation this instability at position $X_{i}$. 
The basic result of the developed theory is the Vlasov equation (\ref{41}) which 
determines the mesoscale  evolution of the ion/electron distribution functions resulted 
from the interaction of ions/electrons with the inhomogeneous IC microturbulence. 
This theory predicts the generation of the sheared poloidal convective flow, and of the 
radial compressed flow with radial flow velocity gradient. 
  
The developed hydrodynamic mesoscale theory reveals the radial compressed convective flow 
as the dominant factor in the formation of the steep pedestal density profile with density 
gradient exponentially growing with time. This gradient density growth is limited by the radial 
oscillating with time outflow of the pedestal ions to SOL. The process of the temporal steeping 
and smoothing of the plasma density profile in the pedestal reveals as the 
manifestation of the ultimate stage of the relaxation of the unstable radially 
inhomogeneous density of the pedestal plasma. It was initiated as the microturbulence 
and finalised as the mesoscale nonlinear relaxation oscillations. The analytical treatment 
of the temporal evolution of the plasma compressed flow near the pedestal bottom at time 
$t > t_{\star}$ may be performed as the solution of the initial-boundary problem 
for the system of equations which includes Eqs. (\ref{81}) - (\ref{84})  for 
$n_{i1}\left(\check{X}_{i}, \tau_{i}\right)$, $T_{i1}\left(\check{X}_{i}, \tau_{i}\right)$, 
$u_{ix}\left(\check{X}_{i}, \tau_{i}\right)$, $u_{iy}\left(\check{X}_{i}, \tau_{i}\right)$,
Eqs. (\ref{81}) - (\ref{84}) for the electron component for 
$n_{e1}\left(\check{X}_{e}, \tau_{e}\right)$, $T_{e1}\left(\check{X}_{e}, \tau_{e}\right)$, 
$u_{ex}\left(\check{X}_{e}, \tau_{e}\right)$, $u_{ey}\left(\check{X}_{e}, \tau_{e}\right)$,
and Eq. (\ref{85}) for potential $\Phi_{i}\left(\check{X}_{i}, \tau_{i}\right)$. 
The solution of this problem will be done in the separate paper.

\begin{acknowledgments}
This work was supported by National R\&D Program through the National Research Foundation of 
Korea (NRF) funded by the Ministry of Education, Science and Technology (Grant No. 
NRF-2018R1D1A3B07051247) and BK21 FOUR, the Creative Human Resource Education and Research 
Programs for ICT Convergence in the 4th Industrial Revolution.
\end{acknowledgments}

\bigskip
{\bf DATA AVAILABILITY}

\bigskip
The data that support the findings of this study are available from the corresponding author upon 
reasonable request.

\appendix 
\section {Velocities $\bar{U}^{(0)}_{ix}\left(\tilde{X}_{i}, t\right)$ and $\bar{U}^{(0)}_{iy}\left(\tilde{X}_{i}, t\right)$}
\label{sec6} 

For the electric field $\tilde{\mathbf{E}}_{i}$, determined by Eq. (\ref{16}), velocities 
$\bar{U}^{(0)}_{ix}\left(\tilde{X}_{i}, t\right)$ and $\bar{U}^{(0)}_{iy}
\left(\tilde{X}_{i}, t\right)$ were determined in Ref.\cite{Mikhailenko}. For completeness, 
we present these results here, i. e.
\begin{widetext}
\begin{eqnarray}
&\displaystyle 
\bar{U}^{(0)}_{ix}\left(\tilde{X}_{i}\right)= \frac{1}{2\omega_{ci}}\frac{e_{i}^{2}}{m^{2}_{i}}
\int d\mathbf{k}\left[a_{i1}\left(\mathbf{k}\right) \tilde{E}_{ix}
\left(\mathbf{k}, \tilde{X}_{i}\right)
\frac{\partial}{\partial \tilde{X}_{i}}\left(\tilde{E}^{\ast}_{iy}
\left(\mathbf{k}, \tilde{X}_{i}
\right)\right)\right.
\nonumber\\ 
& \displaystyle
\left.+a_{i2}\left(\mathbf{k}\right)\tilde{E}_{iy}\left(\mathbf{k}, \tilde{X}_{i}\right)
\frac{\partial}{\partial \tilde{X}_{i}}
\left(\tilde{E}^{\ast}_{ix}\left(\mathbf{k},\tilde{X}_{i}\right)\right)\right]
\nonumber\\ 
& \displaystyle
= \frac{1}{4\omega_{ci}}\frac{e_{i}^{2}}{m^{2}_{i}}\int d\mathbf{k}k_{x}k_{y}\left(a_{i1}
\left(\mathbf{k}\right) +a_{i2}\left(\mathbf{k}\right) \right)
\frac{\partial}{\partial \tilde{X}_{i}}\left|\varphi\left(\mathbf{k}, \tilde{X}_{i}\right)
\right|^{2},
\label{95}
\end{eqnarray}
and
\begin{eqnarray}
&\displaystyle 
\bar{U}^{(0)}_{iy}\left(\tilde{X}_{i}\right)= -\frac{1}{4\omega_{ci}}\frac{e_{i}^{2}}
{m^{2}_{i}}\int d\mathbf{k}\left[ a_{i1}\left(\mathbf{k}\right)\frac{\partial}
{\partial \tilde{X}_{i}}\left|\tilde{E}_{ix}\left(\mathbf{k}, \tilde{X}_{i}\right)\right|^{2}
-a_{i2}\left(\mathbf{k}\right)\frac{\partial}{\partial X_{i}}
\left|\tilde{E}_{iy}\left(\mathbf{k},\tilde{X}_{i}\right)\right|^{2}\right]
\nonumber\\ 
& \displaystyle
= -\frac{1}{4\omega_{ci}}\frac{e_{i}^{2}}{m^{2}_{i}}\int d\mathbf{k}\left(k^{2}_{x}a_{i1}
\left(\mathbf{k}\right)-k^{2}_{y}a_{i2}\left(\mathbf{k}\right) 
\right)\frac{\partial}{\partial \tilde{X}_{i}}\left|\varphi\left(\mathbf{k}, 
\tilde{X}_{i}\right)\right|^{2},
\label{96}
\end{eqnarray}
where the asterisk in Eq. (\ref{95}) implies the operation of complex conjugate. The 
coefficients $a_{i1}\left(\mathbf{k}\right)$ 
and $a_{i2}\left(\mathbf{k}\right)$ are determined as
\begin{eqnarray}
&\displaystyle a_{i1}\left(\mathbf{k}\right)=\left[\frac{\omega_{ci}}
{\omega\left(\mathbf{k}\right)\left(\omega_{ci}+\omega\left(\mathbf{k}\right)
\right)^{2}} +\frac{\omega_{ci}}{\omega\left(\mathbf{k}
\right)\left(\omega_{ci}-\omega\left(\mathbf{k}\right)\right)^{2}}+
\frac{1}{\left(\omega^{2}_{ci}-\omega^{2}\left(\mathbf{k}\right)\right)}\right], 
\label{97}
\end{eqnarray}
and
\begin{eqnarray}
&\displaystyle a_{i2}\left(\mathbf{k}\right)=\left[\frac{1}{\left(\omega_{ci}
+\omega\left(\mathbf{k}
\right)\right)^{2}} +\frac{1}{\left(\omega_{ci}-\omega
\left(\mathbf{k}\right)
\right)^{2}}+\frac{1}{\left(\omega^{2}_{ci}-\omega^{2}\left(\mathbf{k}\right)\right)}\right].
\label{98}
\end{eqnarray}
\end{widetext}
The velocities of electrons $\bar{U}^{(0)}_{ex}\left(\tilde{X}_{i}\right)$ 
and $\bar{U}^{(0)}_{ey}\left(\tilde{X}_{i}\right)$ in the ion frame, with  
electric field $\mathbf{E}_{e}\left(\mathbf{\hat{r}}_{i}, \tilde{X}_{i}, t\right)$ 
determined by Eq. (\ref{17}), are
\begin{eqnarray}
&\displaystyle \bar{U}^{(0)}_{ex}\left(\tilde{X}_{i}\right)\approx \frac{c^{2}}{B^{2}_{0}}
\int d\mathbf{k}\tilde{E}_{ix}\left(\mathbf{k}, \tilde{X}_{i}\right)\frac{\partial}
{\partial \tilde{X}_{i}}\left(\tilde{E}^{\ast}_{iy}\left(\mathbf{k}, 
\tilde{X}_{i}\right)\right)
\nonumber\\ 
& \displaystyle
\times\sum\limits_{p=-\infty}^{\infty}J^{2}_{p}\left(a_{ie}\left(\mathbf{k}, 
\tilde{X}_{i}\right)\right)\frac{1}{\Omega_{p}\left(\mathbf{k}, \tilde{X}_{i}\right)},
\label{99}
\end{eqnarray}
and 
\begin{eqnarray}
&\displaystyle \bar{U}^{(0)}_{ey}\left(\tilde{X}_{i}\right)\approx 
-\frac{1}{2}\frac{c^{2}}{B^{2}_{0}}
\int d\mathbf{k}\frac{\partial}{\partial \tilde{X}_{i}}\left|\tilde{E}_{ix}
\left(\mathbf{k}, \tilde{X}_{i}\right)\right|^{2}
\nonumber\\ 
& \displaystyle
\times\sum\limits_{p=-\infty}^{\infty}J^{2}_{p}\left(a_{ie}\left(\mathbf{k}, 
\tilde{X}_{i}\right)
\right)\frac{1}{\Omega_{p}\left(\mathbf{k}, \tilde{X}_{i}\right)},
\label{100}
\end{eqnarray}
where limit $|\omega_{ce}|\gg \Omega_{p}\sim \omega_{ci}$ was used.

\end{document}